\documentclass{aastex63}

\usepackage{amsmath}
\usepackage{amssymb}
\usepackage{bm}
\usepackage[utf8]{inputenc} 
\usepackage{latexsym}
\usepackage{hyperref}
\usepackage{graphicx}
\usepackage{url}
\usepackage{xcolor,colortbl}
\usepackage{array}

\definecolor{LightCyan}{rgb}{0.88,1,1}
\definecolor{newcolor}{rgb}{.8,.349,.1}

\makeatletter
\DeclareRobustCommand{\cev}[1]{%
  {\mathpalette\do@cev{#1}}%
}
\newcommand{\do@cev}[2]{%
  \vbox{\offinterlineskip
    \sbox\z@{$\m@th#1 x$}%
    \ialign{##\cr
      \hidewidth\reflectbox{$\m@th#1\vec{}\mkern4mu$}\hidewidth\cr
      \noalign{\kern-\ht\z@}
      $\m@th#1#2$\cr
    }%
  }%
}
\makeatother

\submitjournal{ApJS}

\shorttitle{Time domain}
\shortauthors{Solache et al.}

\graphicspath{./}

\begin{document}

\title{Time-domain deep learning filtering of structured atmospheric noise for ground-based millimeter astronomy.}

\correspondingauthor{Iván Rodríguez-Montoya}
\email{irodriguez@inaoep.mx}

\author{Alejandra Rocha-Solache}
\affiliation{Instituto Nacional de Astrofísica, Óptica y Electrónica. Apdo. Post. 51 y 216, 72000. Puebla Pue., México}

\author{Iván Rodríguez-Montoya}
\affiliation{Instituto Nacional de Astrofísica, Óptica y Electrónica. Apdo. Post. 51 y 216, 72000. Puebla Pue., México}

\author{David Sánchez-Argüelles}
\affiliation{Instituto Nacional de Astrofísica, Óptica y Electrónica. Apdo. Post. 51 y 216, 72000. Puebla Pue., México}

\author{Itziar Aretxaga}
\affiliation{Instituto Nacional de Astrofísica, Óptica y Electrónica. Apdo. Post. 51 y 216, 72000. Puebla Pue., México}

\begin{abstract}
The complex physics involved in atmospheric turbulence makes it 
very difficult for ground-based astronomy to build accurate scintillation models and develop efficient methodologies to remove this highly structured noise from valuable astronomical observations.
We argue that a Deep Learning approach can bring a significant advance
to treat this problem because of deep neural networks' inherent ability
to abstract non-linear patterns over a broad scale range.
We propose an architecture composed of long-short term memory cells
and an incremental training strategy inspired by transfer and curriculum learning.
We develop a scintillation model and employ an empirical method to generate a vast catalog of atmospheric noise realizations and train the network with representative data.
We face two complexity axes: the signal-to-noise ratio (SNR) and the degree of structure in the noise.
Hence, we train our recurrent network to recognize simulated astrophysical point-like sources embedded in three structured noise levels, with a raw-data SNR ranging from 3 to 0.1.
We find that a slow and repetitive increase in complexity is crucial during training to obtain a robust and stable learning rate that can transfer information through different data contexts.
We probe our recurrent model with synthetic observational data, designing alongside a calibration methodology for flux measurements.
Furthermore, we implement a traditional matched filtering (MF) to compare its performance with our neural network, finding that our final trained network can successfully clean structured noise and significantly enhance the SNR compared to raw data and in a more robust way than traditional MF.
\end{abstract}


\section{Introduction}
\label{sec:intro}

Millimeter (mm) waves offer a unique astronomical window to study very far and dusty regions of the Universe.
Coincidentally, mm-waves suffer from absorption and re-emission of the water vapor molecules composing the Earth's atmosphere.
At mm-waves, there are $z\lesssim 6$ confirmed dusty star-forming galaxies, and it is possible to observe farther sources if dust is present \citep{Blain2002,Casey2014,Zavala2018}.
Meanwhile, atmospheric turbulence distorts these far-traveling electromagnetic waves along their path through the Earth's troposphere ($\lesssim$10 km), before reaching their final ground-based telescope destination.
Nowadays, large-diameter telescopes and large-format continuum cameras are operational and on course to produce sky surveys with an unprecedented extent and depth \citep{SCUBA2013,Bryan2018,Montana,Brien2018}. However, the atmospheric turbulence still represents a critical challenge for any ground-based mm-astronomy campaign.
The present situation raises the question of whether recent advances in deep learning can provide an attractive approach to deal with the highly structured atmospheric noise recorded in mm-wave continuum observations.

In the past two decades, observations in the far infrared (IR) wavelengths have attracted increasingly more interest from different astronomy fields, probably due to two main particularities:
Due to the abundance of dust (especially in rich star-forming regions) in the Universe, UV and visible light are likely to be absorbed by dust and re-emitted at infrared and mm-wavelengths; nearly half the luminosity radiated from astrophysical objects in the Universe is only accessible at mm-wave windows.
Second, due to an observational effect known as the negative k-correction, the expansion of the Universe does not dilute the thermal-dust energy flux observed at mm and sub-mm waves; consequently, very high-redshift galaxies can be accessible to mm-astronomy with equal ease as lower-redshift sources.
These two unique advantages make the mm-wave window an exciting opportunity to study the star formation history and the evolution of large-scale structures of the Universe.

To continue the far mm-Universe exploration, we need to count on large-area, deep surveys, compounding a representative sample of bright, dim, and high-redshift sources.
Very large-area surveys assembled by space-based missions like Spitzer, Planck, and Herschel \citep{Aghanim2015Planck,Montier2016PlanckHighz,Martinache2018SPHerIC} compose a vast catalog of dusty star-forming galaxies that are being followed-up by spectroscopic measurements; however, they lack the resolution to overcome source multiplicity, and, by design, they could only select the brightest sources; thereby, biasing population estimations.
On the other hand, although ground-based telescopes have much better sensitivity and resolution, they are limited by comparatively lower mapping speeds than space-based telescopes.
Therefore, it is imperative to count on large-aperture telescopes, which, coupled to multi-detector instruments, can realize large-area, high-sensitivity sky-maps in reduced scanning time.

The Large Millimeter Telescope (LMT) contributes to this effort \citep{LMT}, which is the largest mm-wave single-dish telescope, with an aperture diameter of 50 m, and a field of view of 4 \texttt{arcmins}.
At 4600 m above the sea level, on the summit of Sierra Negra Volcano in Mexico, the dry weather allows the LMT to perform high-quality astronomical observations, with winter 220 GHz opacities low as $\tau$=0.025 at wavelengths $\sim$1--3 mm.
These excellent climate conditions are particularly necessary for infrared and mm-astronomy to minimize the water molecules absorption and anomalous emission effects.
One of the LMT's key-instruments has been the Aztronomical Thermal Emission Camera (AzTEC) \citep{wilson2008aztec}, an array of 144 silicon nitride micromesh bolometers distributed into a hexagonal close-packed geometry, designed to perform continuum measurements in the 1.1 mm atmospheric window.
With an average mapping speed of 20 \texttt{arcmin}$^2$ mJy$^{-2}$ h$^{-1}$ (under favorable climate conditions), during the years that it was operational, AzTEC was capable of producing maps with $\sim$500 \texttt{arcmin}$^2$ angular size and 0.2 mJy per-beam depth, in $\sim$50 hours total integration time.

A new generation of mm-wave continuum cameras is currently scheduled to be installed on the LMT, with significantly improved technologies, such as lumped-element kinetic inductance detectors (LEKIDs) \citep{Austermann2018kids,Castillo2018muscat}.
The first camera is the Mexico-UK Sub-millimeter Camera for Astronomy (MUSCAT), a large-format mm-wave camera made of 1600 horn-coupled LEKIDs cooled to an operating temperature of 100 mK \citep{Castillo2018muscat}.
This focal-plane array is designed to observe in the 1.1 mm bandwidth and cover the full four \texttt{arcmins} LMT's field of view.
With a design mapping speed of 3 deg$^2$ mJy$^{-2}$ h$^{-1}$, MUSCAT on the LMT would enable us to map in about 40 h a 160 deg$^2$ region, with a six arcsec resolution, and one mJy depth.
The second camera is TolTEC, a large-format mm-wave imaging-polarimeter made of $\sim$7000 high-responsivity LEKIDs, arranged in three focal planes for the 1.1, 1.4, and 2.0 mm bandwidths, and cooled to a 150 mK operating temperature \citep{Denigris2020cryogenic}.
With 1.9 deg$^2$ mJy$^{-2}$ h$^{-1}$ as the design mapping speed for the 1.1 mm bandwidth, TolTEC would require about 40 h to produce a 100 deg$^2$ multi-color map, with polarization information, 5 arcsecs resolution, and 1 mJy depth (at 1.1 mm).

These advanced instruments will produce valuable astronomical surveys whose raw data will yet be dominated by atmospheric emission.
The first complication with the atmosphere is that it produces highly-structured noise that, in some cases, is up to four orders of magnitude brighter than the astronomical emission of interest.
The second complication, at least for infrared and mm-waves, is a model scarcity in the literature ---presumably due to the complex physics involved--- that could be useful to simulate atmospheric noise in time-domain.
The typical approach to clean atmospheric noise is to compute templates of the common-mode signals seen by all the detectors and then subtract them from the original time-streams \citep{Kovacs2008crush,Sayers2010}.
An equivalent approach is to perform Principal Component Analysis (PCA) to remove the common-mode signals, the standard procedure in the AzTEC pipeline \citep{Scott2008pipeline}.
However, this procedure could be problematic because, in some cases, it can remove astrophysical information too.
Intrinsically, the problem demands an adequate treatment of randomly-varying correlation lengths that cannot be easily handled by classical methods.

Aiming to solve this problem, some of us implemented Independent Component Analysis in map-domain \citep{Rodriguez2018}, a methodology that utilizes high-order statistical moments (negentropy) to exploit the richness of atmospheric noise structure.
However, it would be better to prevent (as much as possible) any contamination leakage to the map-domain in the first place; thus, the cleaning step is more suitable at earlier stages in the domain of time.
In this line of thinking, we see that the challenge of structured noise in time-series does not pertain exclusively to mm-astronomy; other fields have addressed similar issues.
For example, the problem of noisy speech recognition has been addressed by various machine learning strategies (see \citet{Zhang2018review} for a comprehensive review).
Among a few examples in astrophysics, neural network applications have been useful in optical astronomy to correct atmospheric-turbulence distorted images \citep{Suarez2019experience},
variable stars classification \citep{Jamal2020NN-VS},
and have even succeeded in recovering gravitational waves from structured noisy temporal series \citep{George2018,marulanda2020deep}.

In this paper, we propose a deep-learning computational scheme to characterize the expected astronomical and atmospheric signals in time-domain.
Specifically, we propose a structure made of long-short term memory units that can learn to identify astrophysical sources embedded in time-series dominated by structured noise.
To generate diverse sets of training data, we need to develop a physical model of turbulence to reproduce synthetic scintillation data.
Moreover, we utilize real data taken with AzTEC to generate Mock Observations that are more representative of the kind of data to be produced by new generation instrumentation.
Given the physical and statistical complexity involved in this application and the inherently high computational demands, in this paper we aim to present a proof of our concept by processing a single detector signal. Meanwhile, the number of detectors scalability and map-making procedures are deferred to a subsequent paper.

The following section introduces the main aspects of ground-based mm-astronomy observations, which helps us to contextualize the statistical properties of the time-streams used in our study.
In section \ref{sec:deep_learning}, we review the deep-learning theory used to describe the building blocks of our proposed recurrent model.
A more technical section \ref{sec:numerical_exp} is employed to detail our data preparation procedures, hyper-parameter selection, and training strategies.
We discuss most of our quantitative results in section \ref{sec:results}.
Finally, section \ref{sec:conclusion} summarizes our conclusions.

\section{Properties of millimeter wavelength continuum observations}
\label{sec:properties}


In mm-continuum cameras, the detector's array and optical coupling are designed to match the angular size of the telescope beam. Hence, the full array does not entirely sample the sky brightness distribution at any given time. Consequently, a scanning pattern is executed by the telescope to perform a sequential integration recorded by all the detectors in temporal series. Each detector records a stream of time-ordered data that can be modeled as,
\begin{equation}
d(t) = P(t)\cdot s(t) + A(t) + N(t),
\end{equation}
where $s$ is the astronomical brightness distribution,
$P$ the telescope pointing matrix, $A$ the atmosphere emission, 
and $N$ is the instrumental noise.
In the AzTEC architecture, each bolometer couples to a negative temperature coefficient thermistor; hence, it records the sky brightness as a power decrease across the thermistor.
The instrument electronics front-end measures the power signals across the thermistor, digitizes them at a 64 Hz sampling rate, and stores the time-streams in electrical power units (nW).
These values are converted later into radiation flux units,
\footnote{1 Jy $=$10$^{-26}$ W m$^{-2}$ Hz$^{-1}$}
using observations of calibration sources (planets and asteroids).

The data reduction process is performed offline (after the observation time); it usually includes removing large spikes induced by cosmic rays, followed by an atmosphere cleaning process. The pointing calibration data is then used to construct the pointing matrix and project the time-streams into a map. Thus, every pixel in the final map comprises all the detector's temporal sampling average. This clean, reduced map is ready for astrophysical investigations.
For an overview of data-reduction techniques used in mm-astronomy, see Ref. \citep{Mairs2015Comparison}, and for a complete description of the AzTEC pipeline, see \citep{Scott2008pipeline,Rodriguez2018}.

In the following analysis, we consider only \textit{point-like sources} as our astrophysical target.
A point-like source is an astrophysical object with an angular extension significantly smaller than the telescope resolution; hence, in map-domain, they are imprinted as compact sources with the telescope point-spread function's morphology.

Considering typical observation parameters for AzTEC during the 2018 observation campaign (a scanning speed tuned to 50 arcsec s$^{-1}$ and an aperture diameter of 30 m), it leads to a point-like source maximum temporal width of $\sim$4.4 s; which we will use in section \ref{sec:data} as the expected signal width to generate our synthetic data.

We can address two general types of mm-observations: The most frequent are pointing calibrations; these are typically two minutes long integrations on bright point-like sources (a few Jy's) used to calibrate the telescope position accuracy.
The other type is science observations, consisting of integrations that can last from few to hundreds of minutes.
In most cases, science objects are much fainter than the well-known calibration sources.
However, clean and quick pointing calibrations are crucial to ensure the quality of scientific data and optimize the telescope's time.
A successful deep learning strategy can potentially help to improve both the pointing-calibration efficiency and science observations.
Although an intensive computational effort needs to be invested in the training stages, the employed time for inferences is much shorter.

\subsection{Physical modeling of scintillation noise in millimeter astronomy}
\label{sec:structured}

To obtain theoretical predictions for the atmospheric noise typically recorded by continuum cameras, in \ref{sec:propagation}, we develop the physics of electromagnetic waves propagating through the turbulent atmosphere.
We are particularly interested in the random variations of the log-amplitude $\chi\equiv\log(E_1/E_0)$, where $E_1$ represents a small fluctuation around the electric mean-field value $E_0$; which in turn relates to the observed intensity as $(I-I_0)/I_0 = |E_1|^2/|E_0|^2 = e^{2\chi}$ \citep{Tatarskii2016wave,Tatarskii1971}.
We can see that $\chi$ is very useful in quantifying the intensity variations induced by the turbulent atmosphere, also known as \textit{scintillation}.

In the wet atmosphere, mm-waves' scintillation is caused by refraction and absorption turbulent fluctuations, which are sourced by temperature, humidity, and pressure fluctuations.
For infrared and mm-wavelengths, refraction and absorption processes are respectively effectively captured by the real ($n_R$) and imaginary ($n_I$) parts of the refractive index.
Consequently, the scintillation spectral density, $P_\chi(\omega)$, defined by
\begin{eqnarray}
 \overline{\left\langle\chi^2\right\rangle}=\int d\omega \,P_\chi(\omega),
\end{eqnarray}
is given by $P_\chi(\omega) \approx P_R(\omega)+P_I(\omega)$,
where $P_R$ is the contribution from $n_R$ and $P_I$ corresponds to $n_I$.
Here $\omega$ is the Fourier temporal frequency.
The cross-correlated spectral density $P_{IR}$ appearing in Eq. (\ref{eq:app_specs}) is typically much smaller than the previous two, and so, we neglect it.
Considering a telescope with a circular-aperture radius $a_r$, observing at zenith angle $\vartheta$, the explicit expressions are
\begin{eqnarray}
 P_R(\omega) &=& 
  2(2\pi k)^2\sec\vartheta
  \int_{h_{\rm min}}^{h_{\rm max}}dz\,C^2_R(z)
  \int_{\omega/\upsilon(z)}^\infty d\kappa\,
  \frac{\kappa\,\Phi_0(\kappa)}{\sqrt{(\kappa\upsilon)^2-\omega^2}}
  \sin^2\left(\frac{\kappa^2 z\sec \vartheta}{2k}\right)
  \left(\frac{2J_1(\kappa a_r)}{\kappa a_r}\right)^2 , \label{eq:spectrumR} \\
   P_I(\omega) &=&
  2(2\pi k)^2\sec\vartheta
  \int_{h_{\rm min}}^{h_{\rm max}}dz\,C^2_I(z)
  \int_{\omega/\upsilon(z)}^\infty d\kappa\,
  \frac{\kappa\,\Phi_0(\kappa)}{\sqrt{(\kappa\upsilon)^2-\omega^2}}
  \cos^2\left(\frac{\kappa^2 z\sec \vartheta}{2k}\right)
  \left(\frac{2J_1(\kappa a_r)}{\kappa a_r}\right)^2 ,
  \label{eq:spectrumI}
\end{eqnarray}
where $k=2\pi/\lambda$ is the electromagnetic wave-number, and $\kappa=2\pi/l$ is the wave-number associated with the atmosphere fluctuations.
The functions $C_R$ and $C_I$ are the structure parameters related respectively to $n_R$ and $n_I$; $\Phi_0(\kappa)$ is the turbulence spectrum as in Eq. (\ref{eq:Pope}); $\upsilon$ is the transverse wind speed; $J_1$ is the first-order spherical Bessel function.
The integration variable is the altitude, and it goes from the Observatory altitude above sea level $h_{\rm min}$ to the troposphere's end $h_{\rm max}\approx$10 km.

The qualitative behavior of each spectrum is as follows.
The scintillation spectrum due to the real part of the refractive index behaves as a $\omega^{-8/3}$ power-law for high frequencies, but the power depicts a plateau over lower frequencies (see Fig. \ref{fig:theor_scintillation}).
Experimentally, $C^2_I$ has been found to be from $10^{-3}$ to $10^{-7}$ times smaller than $C^2_R$ (see \ref{sec:propagation}).
One might have naively implied that $P_I$ ought to be negligible compared to $P_R$, but that is only true for higher frequencies. Indeed, the imaginary part dominates the lower-frequency power.
It is essential to understand the physics involved in these two frequency domains, especially to plan technological strategies.
For instance, a more sensitive camera with lower thermal noise or a larger telescope with higher aperture efficiency will significantly reduce the high-frequency scintillation noise.
However, none of these two technological improvements would dramatically reduce the low-frequency scintillation noise caused by mm-waves' anomalous dispersion.

Early attempts to introduce a model for the atmospheric-noise temporal power spectrum \citep{Church1995,Lay1997temporal} included explicitly only the refractive-index real-part fluctuations (equivalent to $P_R$), missing the most dominant structure caused by water molecules, and encoded in the refractive-index imaginary part $P_I$.
A useful advantage of the IR Scintillation spectrum in Eqs. (\ref{eq:spectrumR}) and (\ref{eq:spectrumI}) is that they are profile-generic expressions, \textit{i.e.}, different turbulence models $\Phi_0(\kappa)$, structure parameters $C_R^2(z)$, $C_I^2(z)$, or wind-velocity profiles $\upsilon(z)$ can be tested. 
Consequently, the IR Scintillation spectrum can be ideally adapted by physically-motivated parametric approaches \citep{Errard2015modeling} to study the atmospheric effects on astronomical observations.

\subsection{Simulation of atmospheric temporal series}
\label{sec:mockobs}

If we know the noise spectral density beforehand, we can generate random realizations of structured atmospheric noise.
Our model in Eqs. (\ref{eq:spectrumR}) and (\ref{eq:spectrumI}) allows us to generate simulations based on fundamental physics with the appeal of clearly interpreted parameters that are under control.
However, in real observations, there can be many systematic effects introducing several structured noise levels that are not always under control.
Thus, we need a methodology to regard more representative observational conditions, such as the Site's weather and instrumentation.

White Gaussian noise is the simplest example of unstructured noise, characterized by a constant (or flat) power spectrum, which implies stationary uncorrelated temporal series \citep{Boyat2015review}. 
One consequence is that the white-noise standard deviation (std) remains approximately equal regardless of the temporal location and sample size.
Atmospheric noise, on the other hand, is structured with a characteristic Brownian or $1/f$-like power spectrum \citep{Milotti2002}. 
Structured noise is not necessarily stationary; it implies that the global std in a stream may significantly differ from local values within data patches
\citep[see\textit{e.g.}][]{Kirchgassner2012ModernTimeSeries}.
The simplest way to classify the complexity of $1/f$-like noise is through the negative slope in log-log scale; the steeper the spectrum, the more complex the structure. Astronomical observations in mm-waves contain both structured noise from the atmosphere and white noise from instrumentation.

\begin{figure}
\includegraphics[width=\linewidth]{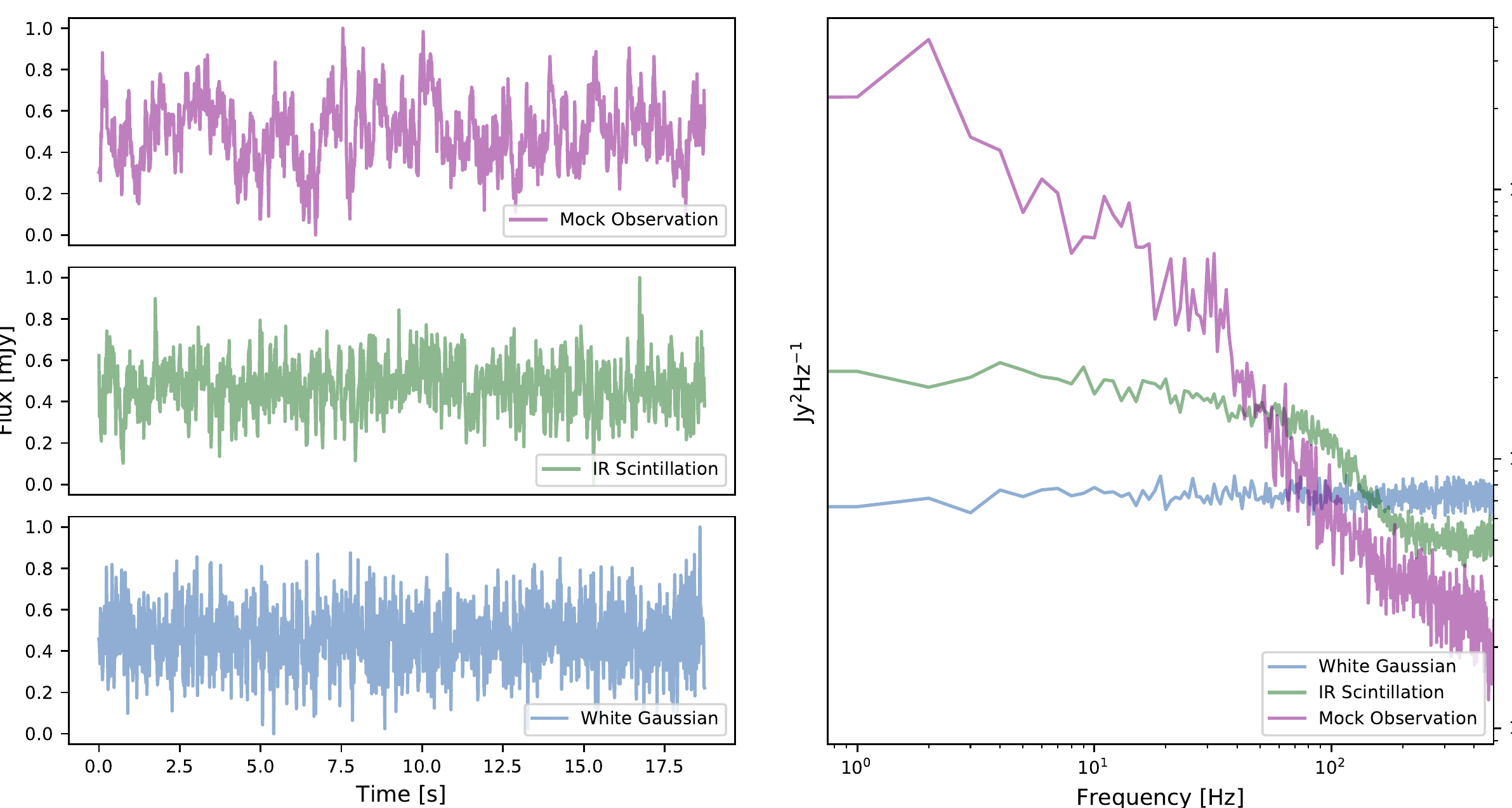}
\caption{
The left panel shows noise-realization examples of White Gaussian, IR Scintillation, and Mock Observations in the time domain. The right panel shows a power-spectra average of 200 noise realizations in the frequency domain.
Mock Observations show higher energy at low frequencies, with an average negative slope near one in log-log space.
White noise resembles a flat spectrum with log-log zero slope.
IR Scintillation noise represents an intermediate complexity used in our training strategy, with log-log slope of approximately -1/2.
See sections \ref{sec:data} and \ref{sec:train} for more details.
}
\label{fig:noises}
\end{figure}

In the presence of structured random noise, it is a common practice to use a whitening filter to enhance the signal detection \citep[see \textit{e.g.}][]{Cuoco_2004,Allen2012,LIGO_2016}.
Conversely, we implement an \textit{inverse-whitening transform} to generate structured-noisy time series, similar to the atmosphere emission at mm-wavelengths.
For the following analysis, we simulate sky observations employing a methodology described by \citet{Rodriguez2018} using a spectral density, either from theoretical considerations or from an observational calibrated data.

Let us assume that $\mathcal{P}$ denotes a target spectral density (derived theoretically or inferred from observations). Likewise, $\mathcal{N}_i$ denotes a white-noise flat power spectrum.
Then, the $i$-th simulated noisy realization $\mathcal {S}_i$ can be generated by
\begin{equation}
 \mathcal{S}_i = \mathcal{F}^{-1}\lbrace (\mathcal{P}+\mathcal{N}_i)^{1/2} \, \mathcal{G}_i \rbrace ,
 \label{eq:atm_sim}
\end{equation}
where, $\mathcal{F}^{-1}$ denotes the inverse-Fourier transform operator and $\mathcal{G}_i$ is a Fourier-transformed random-Gaussian sequence with the same length of the time-stream. 
Effectively, $\mathcal{G}_i$ allows us to model small power shifts from the average power spectrum.
Notice that $\mathcal{S}_i$ follows a Gaussian distribution, with correlation scales that depend on the random combination of $\mathcal{P}$, $\mathcal{N}_i$, and $\mathcal{G}_i$.%
\footnote{
Non-Gaussianities are expected in actual atmospheric data, and thus, the spectral density may not contain all relevant information.
Nonetheless, our scintillation model could be combined with iterative techniques like the Johnson translator system \citep[see \textit{e.g.},][]{Wang2018simulation,Wang2021generating} to simulate atmospheric series with a given skewness and kurtosis.
}

Thus, the power-spectrum shape (either flat or $1/f$-like) determines the degree of structure in our atmospheric noise simulations, and we can generate three levels of structured noise:
\textit{i)} the simplest White Gaussian noise is generated with $\mathcal{P}=0$, 
\textit{ii)} the IR Scintillation noise with $\mathcal{P}$ given by Eqs. (\ref{eq:spectrumR}) and (\ref{eq:spectrumI}) with added white instrumental noise, and \textit{iii)} Mock Observational noise generated from an empirical spectral density.
Rather than just picking a single one, we exploit these three incremental structure levels to train our neural network and improve overall learning performance.
Fig. \ref{fig:noises} shows examples of the three types of simulated structured noisy inputs that we use throughout the neural network training.

\section{Deep learning and recurrent networks}
\label{sec:deep_learning}

\subsection{Neural networks and recurrent connections}
\label{sec:neuralnetworks}

Recurrent Neural Networks (RNN) are a particularly attractive deep learning architecture for applications involving ordered data. We can understand an RNN cell as a function that exploits recursiveness to find the current state of a dynamical variable. Keeping correlations among spaced events is indeed a powerful learning tool to handle long-standing structures.
An RNN cell morphology is similar to the most basic cell but with a self-connected loop –namely, a \textit{hidden state}– that feeds its own output back a number of times, turning the cell into a dynamical system that correlates the information between individual inputs \citep{cho2014learning}.
The simplest example is the Jordan RNN cell \citep{jordan1997serial}, which is composed of two units; the first one is a hidden state representing the memory of the cell,
\begin{eqnarray}
 h_t &=& f_h(U_h\,x_t + W_h\,y_{t-1} + b_h),
 \label{eq:JordanIn}
\end{eqnarray}
where $x_t$ is the current input state, and
$U_h$ is an input-to-hidden weight matrix.
By the same token, $y_{t-1}$ is the previous output state,
$W_h$ is an output-to-hidden weight matrix,
and $b_h$ is a bias parameter.
The function $f_h$ is a non-linear transformation that produces the current hidden state $h_t$.
The second component of the Jordan RNN cell is in itself the most basic unit used in deep learning (also called the \textit{dense} cell), and it produces the output state in this case,
\begin{eqnarray}
  y_t &=& f_y(\,V_y\,h_{t} + b_y),
  \label{eq:JordanOut}
\end{eqnarray}
in which, a hidden-to-output weight matrix $V_y$ multiplies the hidden state $h_t$,
with the addition of a bias parameter $b_y$, and non-linearly transformed by the function $f_y$.
The output state $y_t$ can either be fed back to the cell, or delivered to other cells as their input.
The non-linear transformations $f_h$ and $f_y$ are called \textit{activation functions}, and help to assigns a relevance to the cell's information output.
The RNN cell abstracts features of the temporal sequence in its weight matrices,
where $U_h$ is more sensitive to new external data, and $W_t$ keeps a memory of the previous state.
Since the only purpose of the hidden state $h_t$ is to abstract the sequence's temporal behavior,
the hidden state $h_t$ may have arbitrary dimensions.

As a drawback of recurrent networks, the amount of information that can be abstracted into a simple RNN unit is typical of short-range.
Also known as \textit{vanishing gradient effect} \citep{hochreiter2001gradient}, there is always a risk of inducing learning degradation over extended time intervals.
Closely related is the problem of \textit{exploding gradients} \citep{pascanu2013difficulty}, when cumulative losses increase excessively and lead to unstable or impossible learning.
A better practice is to use the Long Short-Therm Memory (LSTM) \citep{hochreiter1997long} to avoid the vanishing and exploding gradient effects.

The LSTM cell is a discriminating mechanism to either preserve or discard long-term information.
The LSTM cell's ability to forget unnecessary information enables itself to keep temporal correlations for more extended periods.
Intuitively, an LSTM unit internal structure is made of two hidden mechanisms that act as pipelines for the information of distinct temporal scales.
The first mechanism performs the operations of a single recurrent cell, compiling a representation of a few past data inputs.
This hidden state is interpreted as a short-term memory unit, due to its limited ability to abstract small-extent behavior.
The second mechanism encodes the sequence's global behavior, acting as a filter for the most significant long-term features.
This new memory state allows the network to keep important information that would otherwise be overwritten by new data entries in a short-term memory unit.

\begin{figure}
\includegraphics[width=\linewidth]{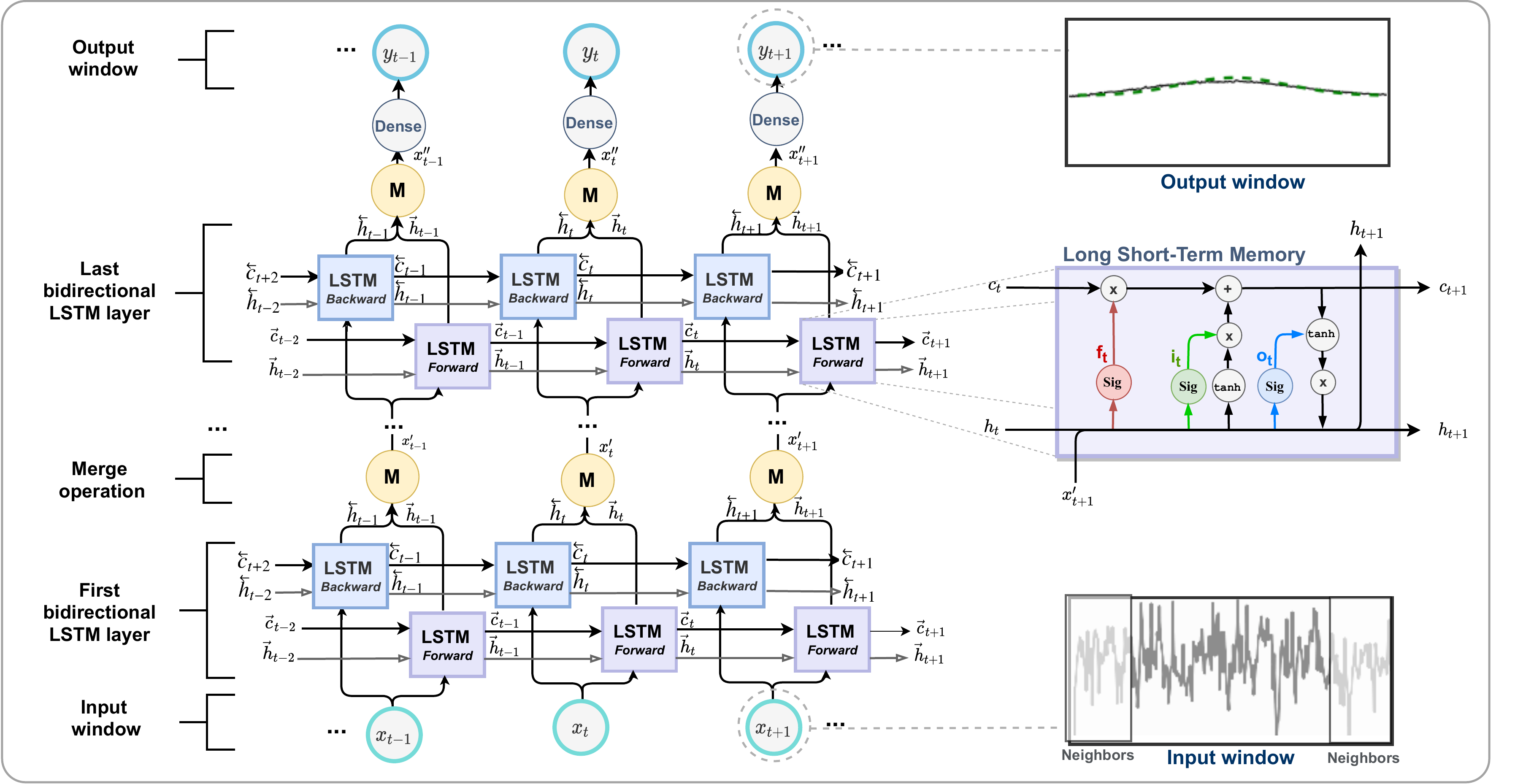}
\caption{
Visual representation of our deep recurrent model, mainly built up of LSTM units deployed in bidirectional layers.
L.H.S: At each training iteration $t$, a noisy input window $x_t$ is processed by the first LSTM forward and backward layers.
The forward LSTM sub-layer learns the temporal structure form left to right, while its twin backward sub-layer learns in the opposite direction.
An intermediate module M performs a merge of forward $\vec{h}_t$ and backward $\cev{h}_t$ hidden states.
This operation yields a new state $x'$ with augmented information to feed the upper bidirectional LSTM layer.
As a final processing step, a linear \textit{dense} layer transforms the last layer output into an object $Y_t$ with the predefined output shape.
On the R.H.S., the LSTM cell internal structure is exposed. The \textit{cell} state $c_t$ or long-term memory, along with the \textit{hidden} state $h_t$ or short-term memory, both preserve an internal representation of distinctive temporal scales.
The information share between these two states is carried out by the \textit{forget} $f_t$, \textit{input} $i_t$, and \textit{output} $o_t$ logic gates. See section \ref{sec:neuralnetworks} for more details.
   }
   \label{fig:requiem}
\end{figure}

In the vanilla LSTM layer \citep{hochreiter1997long}, depicted on the right-hand side of Fig. \ref{fig:requiem}, the information flows through a system of logic gates whose parameters are trained to learn what information should be preserved or missed.
The forget $f_t$ and input $i_t$ gates can be written as,
\begin{eqnarray}
 f_t &=&\text{Sig}(U_f\,x_t + W_f\,h_{t-1} + b_f ) ,  \\ 
 i_t &=& \text{Sig}(U_i\,x_t + W_i\,h_{t-1} + b_i). \nonumber
 \label{eq:gates}
\end{eqnarray}
Notice that each logic gate has a simple recurrent cell structure, with the only distinction of their performed task for the LSTM.
The forget gate $f_t$ uses the Sigmoid activation function to evaluate the data input $x_t$ and the previous hidden state $h_{t-1}$ (along with their respective weight matrices) and returns numerical labels between 0 and 1.
A label close to zero means that a specific memory entry is not relevant for the global understanding of the sequence, and hence, it will be likely neglected in subsequent iterations.
Conversely, a label close to one means that the state is relevant and is likely to be preserved.
The second task is to evaluate how relevant is the new input of data, which is done by the input gate $i_t$, also using the Sigmoid function and the respective weight matrices.
Both $f_t$ and $i_t$ are then combined to produce a new long-term memory called the \textit{cell state},
\begin{eqnarray}
 c_{t} &=& f_{t}\,\circ\,c_{t-1}+i_t\,\circ\,\text{tanh}(U_c\,x_t + W_c\,h_{t-1}).
 \label{eq:cell_state}
\end{eqnarray}
The Hadamard (or element-wise) product $f_t\circ c_{t-1}$ suppresses the irrelevant features of the past cell state $c_{t-1}$ and prevents them to persist into the new cell state $c_t$.
The $\tanh$ gate is an additional recurrent operation used to complement the input gate $i_t$ to create a list of candidates integrated into the long-term memory $c_t$.
Finally, the output gate has the role of actually selecting from the feature candidates contained in the new cell state $c_t$, those that will be transferred to the new hidden state $h_t$.
This operation is just a filtered version of $c_t$, with a simple Sigmoid gate $o_t$,
\begin{eqnarray}
 o_t &=& \text{Sig}(U_o\,x_t+W_o\,h_{t-1} + b_o),  \label{eq:output_state}  \\
 h_t &=& o_t\,\circ\,\tanh(c_t). \nonumber
\end{eqnarray}
The output state $h_t$ can be either used as a hidden state fed back to the cell, or delivered to the next LSTM layer as an input state.

The fact that each network logic gate can (be trained to) decide what information is more or less relevant (to describe the data of interest) is a remarkable advantage of the LSTM cell.
In the family of recurrent architectures, the LSTM indeed represents state of the art in modeling complex systems that require large scales of temporal understanding.
For a comprehensive explanation of the LSTM cell internal structure and its variations, we refer the reader to \citep{hochreiter1997long,greff2017lstm}.

\subsection{LSTM network for time-domain astronomical cleaning}
\label{sec:model}

Even though the atmosphere evolution is chaotic, inside the universal inertial regime (see section \ref{sec:propagation}), it is possible to find a time-scale in which the fluctuations are coherent and well-described by the spectral density in Eqs. (\ref{eq:spectrumR}).
Consequently, the LSTM cell seems to be a natural choice to discern among atmospheric and astrophysical behaviors.
Below, we describe our recurrent architecture implementation to handle temporal astronomical data.

Given that the astronomical data reduction process is performed offline (\textit{i.e.}, after the observation time), we can modify the LSTM structure to benefit not only from past events but also from future information.
Thus, we choose a many-to-many input/output configuration: a sliding window replaces each time segment in the series with a fixed number of past and future neighbors. 
We feed every temporal window into the network with the aim to abstract the neighboring correlations.

Taking even more advantage of future temporal information, we implement a bidirectional approach \citep{schuster1997bidirectional,graves2005framewise} that combines two identical recurrent units into a single layer: one processing the input series in a right-wise direction, and another processing the same time series left-wise.%
\footnote{During the preliminary stages of our model, we explored a recurrent encoder-decoder configuration \citep{cho2014learning}. Yet, the implied high-computational costs did not pair with a meaningful learning loss value improvement (for our purposes).}
Recalling the notation used in Eqs. (\ref{eq:cell_state}) and (\ref{eq:output_state}), each bidirectional LSTM layer computes two independent cell states $\vec{c}_t$ and $\cev{c}_t$ that encode the long-term memory in the right-wise and left-wise directions, respectively.
Likewise, each bidirectional LSTM layer produces two hidden states $\vec{h}_t$ and $\cev{h}_t$.
In our implementation (see the M modules in Fig. \ref{fig:requiem}), the merge of these two states creates the bidirectional LSTM layer output,
\begin{eqnarray}
 y_t = \text{M}(\vec{h}_t, \cev{h}_t).
\end{eqnarray}
The function M is a concatenation operation that ties the states into a unique output $y_t$,
which becomes in turn the input for the next bidirectional LSTM layer $y_t\rightarrow x'_t$.

According to Fig. \ref{fig:requiem}, the information flows through our architecture as follows. A time-segment of a data stream $x_t$ is fed into the first bidirectional LSTM layer. The forward layer sees the sequence in the causal temporal sense, from left to right. The backward layer processes the window simultaneously in the reversed direction.
The cell states $\vec{c}_{t-1}$, $\cev{c}_{t+1}$, and the hidden states $\vec{h}_{t-1}$, $\cev{h}_{t+1}$ are initialized with the outputs from the preceding and following temporal windows. The forward and backward LSTM cells produce two hidden states $\vec{h}_t$ and $\cev{h}_t$ that are joined by a merge module M and then delivered as the input $x'_t$ to the next bidirectional layer. In this way, the information obtained from both temporal analyzes is packed into a new single matrix representation.
The process continues up to the deepest bidirectional LSTM layer contained in the network. At the last processing stage, there is a simple dense layer (without any recurrent connection, see Eq. \ref{eq:JordanOut}), which acts just as a parametric point-wise operation to map the LSTM output into the user-defined temporal size (we choose this hyper-parameter value in section \ref{sec:hyperp}).

\section{Numerical experiments}
\label{sec:numerical_exp}
In the previous sections, we explained our expected signal's physical and statistical properties, connected to the deep learning theoretical framework for our architecture's design.
In this section, we describe the data preparation to train the network, the choices for hyper-parameter values, and the training strategy that we adopt to produce the results discussed in the next section.

\subsection{Data preparation}
\label{sec:data}

\begin{figure}
\includegraphics[width=\linewidth]{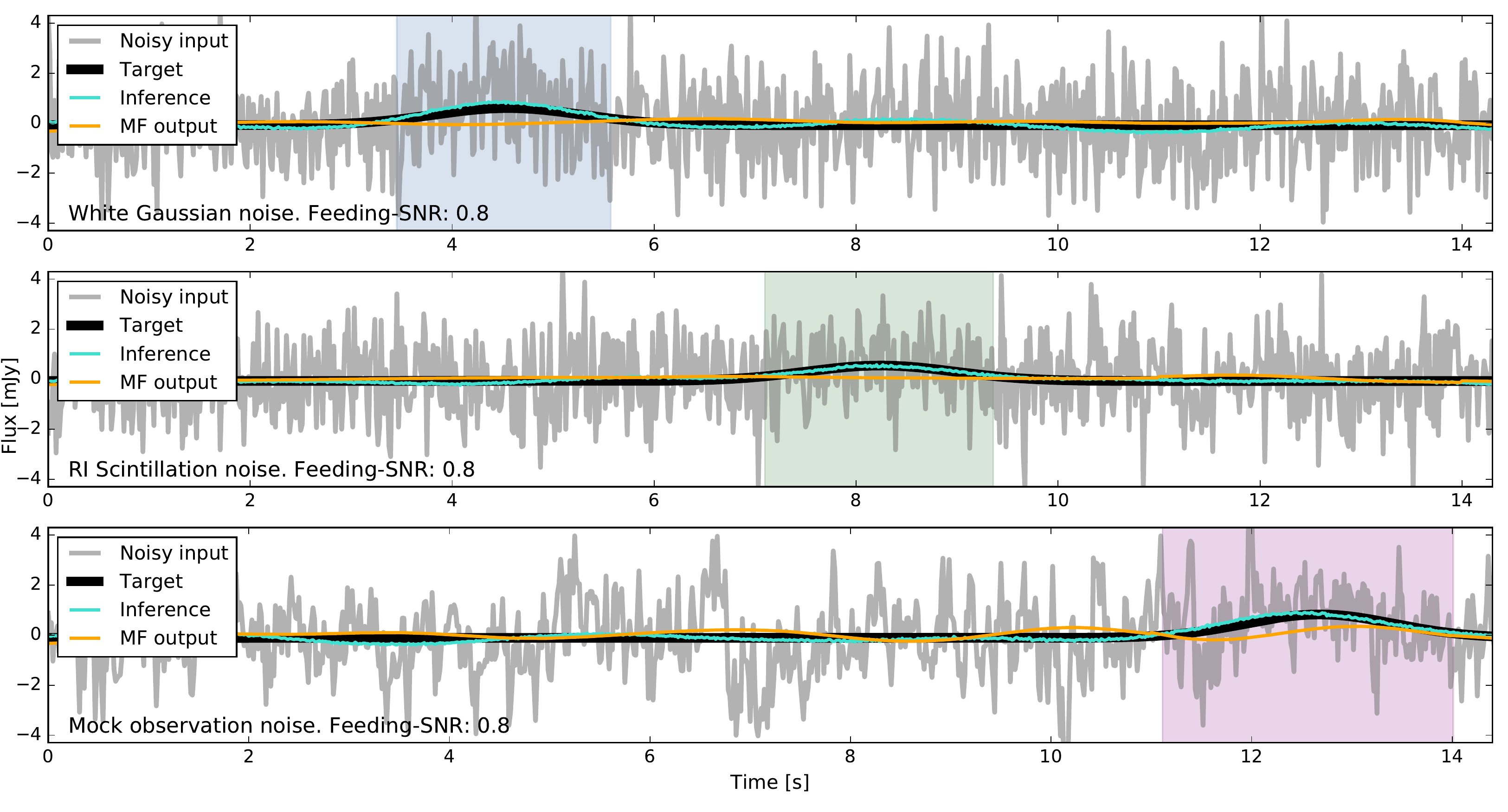}
\caption{Examples of three levels of the structured noise and target signals (see sections \ref{sec:mockobs} and \ref{sec:data} for details on data preparation) filtered by our recurrent neural network.
The gray lines are noisy inputs, the three of them with an initial signal-to-noise ratio of 0.8;
black lines represent a simulated astronomical signal dubbed as the target;
the light turquoise lines are the network inferences of the underlying astronomical signal, which are detected after the
cleaning process with a signal-to-noise ratio above 3.
For comparison, a Matched-Filtered (MF, section \ref{sec:mf}) sequence is shown (orange line), failing to detect any source at this very low feeding-SNR.
(See Fig. \ref{fig:inferences_zoom} for a zoom in the source detection windows.)}

\label{fig:inferences}
\end{figure}

The effectiveness of deep learning algorithms relies on the quality of training: the more massive and accurate the datasets, the higher are the expectations of improvement.
These two conditions are obviously uncommon in many real-world applications and particularly for unexplored astronomical sources, for which the underlying brightness is \textit{a priori} unknown and always mixed with foregrounds and instrumental noise.
Notwithstanding, it is possible to train the network with synthetic data created on the basis of physical models, intending to transfer the gained learning to more realistic applications.
Provided that mock and real data domains are closely related, it has been extensively documented that the knowledge can be transferred from the former to the later domain \citep{pan2009survey}.

We generate temporal series with point-like source random insertions, choosing their statistical properties according to instrumental and observation specifications, as described in section \ref{sec:properties}. The network processes a single-bolometer recording as an independent input.
As Figs.\ref{fig:noises} and \ref{fig:inferences} exemplify, we consider three types of structured noise with increasing complexity:
 \textit{i)} White Gaussian noise is the less structured noise under consideration. Despite its unrealistic simplicity, this noise distribution works as a handy starting point for our training strategy.
\textit{ii)} IR Scintillation noise samples represent a complexity midway point. We choose an intermediate value 10$^{-5}$ for the ratio $P_I/P_R$ in our simulations (see also appendix \ref{sec:propagation}).
\textit{iii)} Mock Observations synthesized after real data are our closest approximations to the observed atmospheric foreground and expected instrumental features.

We perform a few preprocessing steps for the real AzTEC temporal series to make them numerically more treatable.
We utilize 2018 season real AzTEC observations with a 0.17 opacity, representing remarkably bad weather nights. We apply PCA to a 110 time-stream set and then subtract the first two principal components. 
This process removes typically the longest correlations linked mainly to bad weather conditions.%
\footnote{
PCA can partially remove astronomical emission too. Thus, a more accurate preprocessing methodology or a flux calibration \citep[as in][]{Downes2012} could be employed in the future with real-time observations.}
Meanwhile, the remaining noise is still considerably more structured than the IR Scintillation and White Gaussian simulations (see Fig. \ref{fig:noises}).
Finally, we pick a detector located at the edge of the array; because it is off-source most of the time, the atmospheric flux dominates this detector's time-stream, and we can conveniently use it for further simulation steps.
Training the neural network with realistic simulations allows us to build any features from other systematic effects (\textit{e.g.}, the faint confusion background) into the LSTM parameters, helping the network to recognize the embedded targets.

Next, for the three types of noise, we fit and subtract a third-degree polynomial that allows us to work with centered data and filters out the larger scales. We proceed to insert point-like source signals into the simulated noisy time-streams.
Due to long-range non-stationarity in some structured noise samples%
\footnote{
In a series coming from a non-flat power spectrum, we may have for instance that std(15$s$) $\neq$ std(2$s$).},
we should avoid interpreting the global-std as the relevant level of noise.
The time-stream size is chosen from technical limitations such as computational memory or telescope's time, so that there is nothing special about the global std.
Instead, we choose the full point-like source extent (4.4 $s$ at $6\sigma$) to compute the local std as the reference noise.
We adopt the following procedure: we randomly select a time segment that matches the point-like source; then, we measure the noise-std inside 4.4 s and adjust the point-like source amplitude to meet the desired signal-to-noise rate (SNR). We name this induced complexity as \textit{feeding-SNR}, calculated as the maximum signal's amplitude ratio to the local noise level. This procedure allows us to control the complexity presented to the deep-neural network at every training stage.

By construction, for any of the three types of noise, the source brightness relative to the local noise is the only relevant independent variable to classify the sources in complexity. Thus, as a final preprocessing step, we scale the temporal series to have the same standard deviation. It means that our simulations could be interpreted either as bright calibration sources immersed in very noisy data or as dim sources in moderate noise.

We stress that our time-stream simulator feeds the network with unique random realizations of structured noise. This aspect is especially relevant because deep neural networks are vulnerable to overfitting when training is performed over long periods using a finite dataset. In our case, however, we can generate virtually as many random realizations as the training procedure may require, while each noise pattern passes through the network only once.

 \begin{figure}
\includegraphics[width=\linewidth]{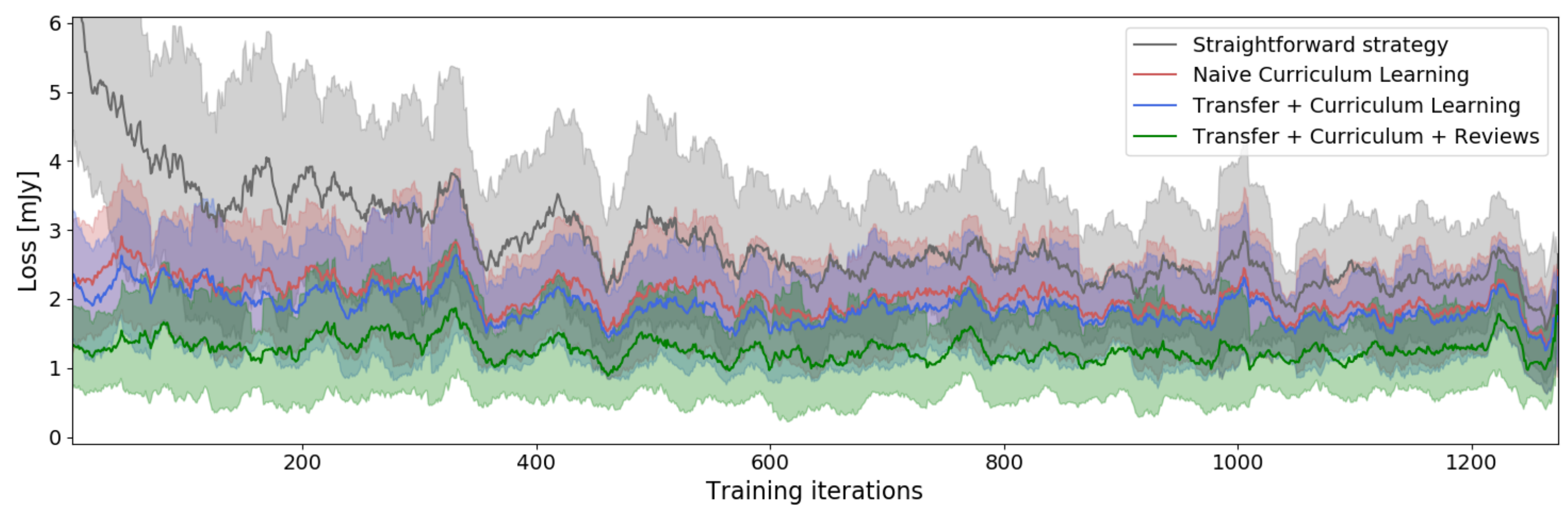}
\caption{
Learning curves along the first 1250 training iterations with mock-observations noise and random feeding-SNR between 0.1 and 3.1. Previously, each model has been thoroughly trained according to the learning strategies drawn in Fig. \ref{fig:paths}.
Each curve is measured as the average MSE evolution, and the colored regions represent the corresponding dispersion. Four training strategies are tested, from which the \textit{straightforward} strategy is the less effective, while our proposed strategy \textit{curriculum transfer learning with spaced reviews} exhibits the best performance. (See section \ref{sec:train} for further discussion.)
}
\label{fig:loss}
\end{figure}

\subsection{Hyper-parameters}
\label{sec:hyperp}

Before designing neural network architecture, we first need to choose the batch dimensions, \textit{i.e.}, the number of time steps feeding the LSTM. Then, the network uses each target stream and its corresponding inference to compute the relative loss function that serves as a reference to update the network weights. We can significantly reduce the memory resources by splitting the batch into a group of windows, which cannot be arbitrarily small. Although point-like source signals are a few seconds long, the LSTM short-term memory needs to abstract longer time scales to assemble the long-term memory structure patterns. Thus, the window size must be at least broader than the point-like source length.  Moreover, individual windows should conveniently share some intermediate time-steps to help the network performing continuous inferences through cutting edges.

Balancing the underlying application requirements and computational costs, we choose a batch size of 1,200 time-steps (equivalent to 18.7 s), containing two 700 time-step-long windows, sharing 100 neighbors on both sides.
It means that for each window, 700 time-steps are fed into the network to infer the mid 500 time-steps. At this point, the batches are ready to feed the network.

A deep neural network design requires selecting a set of hyper-parameters representing the network capabilities and computational constraints: more extensive networks imply more massive processing time and memory.
To avoid the exponential complexity involved in a blind search (\textit{i.e.}, exploring every combination inside the hyper-parameter space grid), some systematic approaches like the random search, Bayesian models, genetic algorithms, among others, can be used to produce efficient results \citep{luo2016review}.
Nonetheless, naive or expert-based explorations have not been entirely suboptimal in a wide variety of applications \citep{bergstra2011algorithms}.

Our approach consists of a semi-systematic exploration to find suitable hyper-parameters values. First, we set up an initial framework with a simple configuration: we employ the well know Adam optimization algorithm \citep{kingma2014adam}, looking to minimize the Mean Square Error (MSE) as the loss function. Then, we perform a two-step hyper-parameters search corresponding to the size of the architecture.
We explore the network's depth (number of layers) and width (number of cells per layer) parameters monitoring the loss function to find the best performance.
The depth dimension is explored from one to seven LSTM layers with increments of one.
We sweep the width dimension in the range from 40 to 250 cells in each layer with increments of 30 cells. The depth-width search space is delimited into a $7\times 9$ grid using a fixed learning rate equal to $10^{-3}$.
Finally, we perform a fine-tuning of the learning framework by iterating over Adam's learning rate range between $10^{-3}$ to $10^{-9}$. As a result, we instantiate the model proposed in section \ref{sec:model} with 190 bidirectional LSTM cells, four deep layers and a $10^{-5}$ value for Adam's learning rate.

\begin{figure}
\centering
\includegraphics[width=0.45\linewidth]{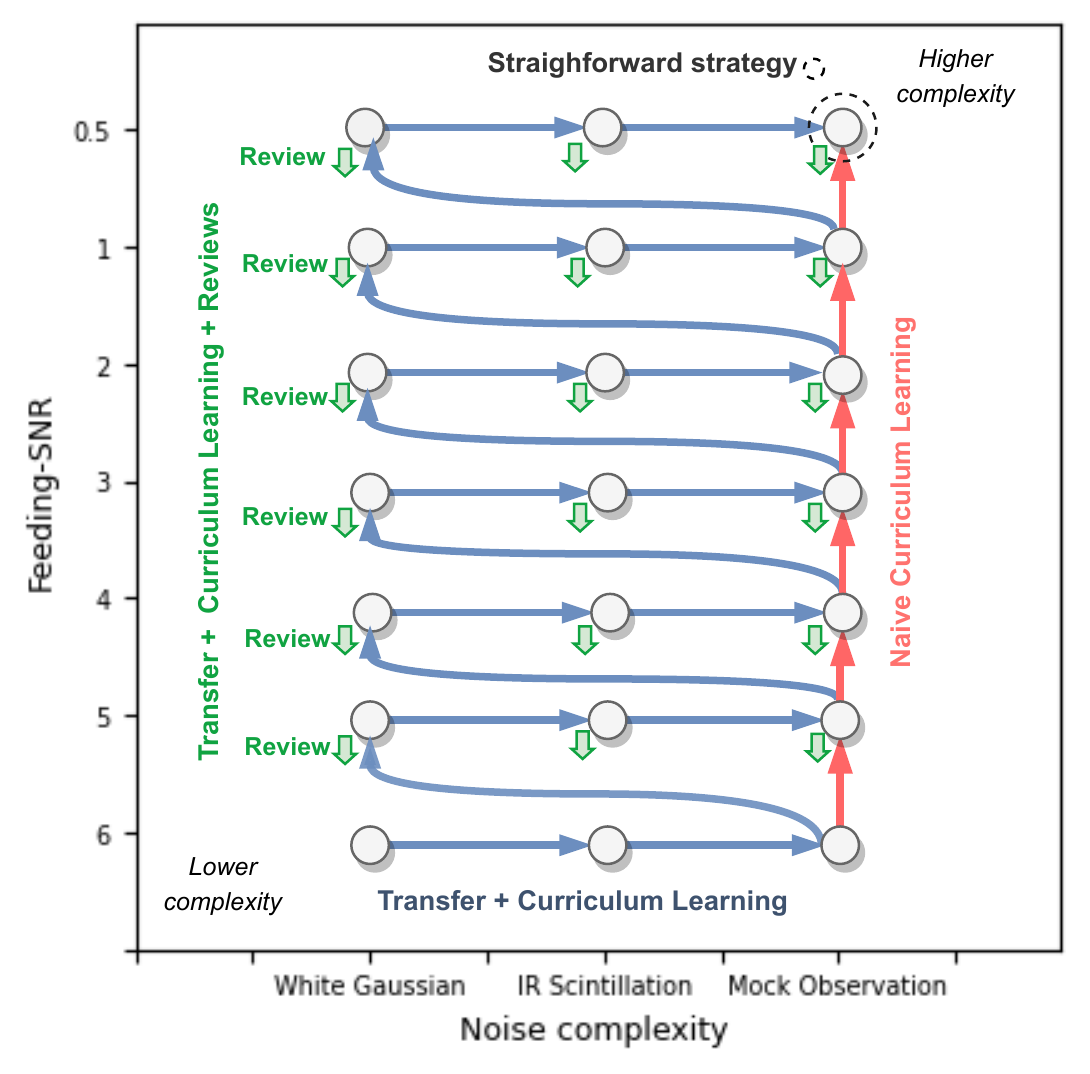}
\caption{
Pictogram of the paths taken by each training strategy tested. It should be read from the left-lower to the right-upper corner. The exploration space is built onto two axes: the feeding-SNR on the vertical axis and the noise complexity level on the horizontal axis. The right-upper dotted circle represents the Straightforward strategy. The red-solid line going upwards at the right corresponds to the Naive Curriculum Learning strategy. The blue-solid zig-zag line represents Transfer + Curriculum Learning. Finally, the green pointing-down arrows denote the Transfer + Curriculum + Reviews strategy. (See section \ref{sec:train} for details.)
}
\label{fig:paths}
\end{figure}

\subsection{Training}
\label{sec:train}

The structure degree reached by the atmospheric noise represents quite a tough starting point to be handled by any parametric architecture, especially for weak-brightness sources.
Nevertheless, a training strategy inspired by curriculum \citep{learningToExecute} and transfer learning \citep{weiss2016survey} is ---as we see below--- able to perform a gradual complexity exploration over several levels of structure and feeding-SNR.

Transfer and curriculum learning are techniques to share knowledge between machine learning models.
On the one hand, curriculum learning aims for intuitive exploration, beginning with simple examples and gradually escalating complexity.
On the other hand, transfer learning takes advantage of models trained in a data domain, using their resulting parameter values (or part of them) as initialization states to find better results in a related but new data domain.

To perform a stable training with the highest attainable precision, we propose an incremental learning approach that combines transfer and curriculum learning to capture two complexity axes: the initial SNR and the type of noise structure.
Dubbing curriculum learning as the strategy's vertical axis, our application gradually decreases the feeding-SNR from 3.5 to 0.1. This procedure helps to accelerate the optimization convergence by simulating the gradual inclusion of fainter astrophysical sources.
Correspondingly, we dub transfer learning as the strategy's horizontal axis: at each feeding-SNR level, the network explores three structure degrees: White Gaussian, IR Scintillation, and Mock Observations. This procedure pretends to mimic the conditions of incrementally worse weather.

We monitor the learning evolution along the training stages regarding the loss function (the target-inference MSE) at each training iteration, which is a good practice for early identification of miss behaviors such as overfitting or deficient learning performance.
Monitoring the loss curve at intermediate stages also allows us to register variations from different training strategies.
For example, the simplest approach would be the \textit{straightforward} strategy, in which the network is trained directly with the highest complexity, in this case, the noisier Mock Observations.

However, we probe three more learning strategies (explained below), each one exploring different routes of incremental complexity into a discretized two-dimensional space. Each training-strategy path's design is illustrated in Fig. \ref{fig:paths}. After a full network training, each strategy is tested on the same dataset of 1250 series with mock-observations noise and a randomly varying feeding-SNR between 0.1 and 3.5. In this test, the same dataset is used to make a fair comparison of the learning strategies.
In Fig. \ref{fig:loss}, we record the loss-function value at each iteration for every tested training strategies. The thick lines represent the loss-function moving average inside a 25-iteration local window, and the colored bands represent the corresponding moving std.

For the straightforward strategy, we can see in Fig. \ref{fig:loss} that the loss-function displays a sudden drop-out at the training beginning ($\lesssim$600 iterations), due to the initial network adaptation to the data domain. A few steps later ($\gtrsim$600 iterations), a plateau can be noticed, albeit with disadvantageous dispersion amounts.
We call the second strategy \textit{naive} curriculum learning (NCL), which consists of a preliminary training over Mock Observations, with feeding-SNR constant decrements from 3.5 to 0.1 in steps of 0.1. At any given feeding-SNR bin, when no significant improvement in the last 400 iterations is noticed, the network proceeds to the next feeding-SNR bin. We notice that NCL considerably improves the loss mean-value compared to the straightforward strategy, albeit the dispersion levels are still comparable. Besides, the NCL training check-point performs poorly with White Gaussian or IR Scintillation noise inferences. This lack of robustness against structure degree variations suggests the main disadvantage of NCL.

Thus, we test a third strategy called \textit{transfer-curriculum learning} (TCL), which alternates transfer learning iterations (using sequentially White Gaussian, IR Scintillation, and Mock Observations) amid intermediate stages of curriculum learning.
TCL reports a slightly lower average loss and less dispersion than NCL, but with the added benefit of performing better with the other two types of noise.

Notice that NCL and TCL loss-curves suffer considerable variability, some of them as large as those reported by the straightforward strategy. This dispersion is rooted in the network's adaptation to the latest stages while forgetting previous feeding-SNR levels, an effect also known as \textit{catastrophic forgetting} \citep{mccloskey1989catastrophic,ratcliff1990connectionist}.
It means that a chronic poor understanding of past training samples might pile up as the relative complexity increases. 
This effect would appear after long periods of training and an over-specialization of the latest feeding-SNR bins.

Inspired by a human learning strategy known as spaced repetition
\footnote{In cognitive psychology and pedagogy, spaced repetition is a learning technique based on the periodic review of concepts to reduce the probability of forgetting information after long studying periods.},
we address the catastrophic forgetting problem, by adding intermediate training stages to \textit{review} the past complexities.
We call this strategy \textit{transfer-curriculum learning with spaced reviews} (TCL-SR).
After each transfer and curriculum learning round (between two feeding-SNR bins), we schedule a review of the previous feeding-SNR range.
By randomly choosing the feeding-SNR, the network can ``remember'' high-SNR data distributions while continuing learning lower SNR complexities.
During every reviewing stage, the iteration number is controlled on-the-fly by the loss-function convergence; 
after 400 iterations without significant MSE improvement, the training scheduler triggers the next TCL-SR stage to pursue the following complexity level.
In comparison, the TCL-SR strategy yields better results in terms of narrower variability and lower mean values for the loss function.
Although the invested training time in the TCL-SR strategy is more significant than the straightforward strategy,
the benefits are appreciated, especially at late stages,
when the low SNR and high structured noise levels pose a more challenging task for network inferences.

Thus, we implement the TCL-SR strategy to train our LSTM-network; 
the training procedure continues until an improvement in the loss-function is no longer noticeable.
Then, the optimization module may be disconnected, and the free parameters get frozen.

\subsection{Matched Filtering}
\label{sec:mf}


We implement a matched-filtering (MF) to compare our trained network results to conventional signal processing techniques. A similar approach has been used as a baseline to compare the performance of neural networks, particularly in gravitational-wave experiments \citep{Cuoco_2004, Allen2012, LIGO_2016}.
The MF is defined in the frequency domain by \citep{Turin1960MF},
\begin{equation}
    \mathcal{M}(\omega)=\frac{\mathcal{H}^\star(\omega)}{\mathcal{P}_N(\omega)}, 
\end{equation}
where $\mathcal{H}$ is a template for the signal of interest (the star superscript denotes complex conjugation), and $\mathcal{P}_N(\omega)$ is the noise-spectral density.
The MF is applied to the raw signal $D(\omega)$ in Fourier space, and the filtered signal is computed from the inverse Fourier transform,
\begin{equation}
    \hat{s}(t) = \Re \left(\mathcal{F}^{-1} \lbrace D(\omega)\star\mathcal{M}(\omega) \rbrace\right).
\end{equation}
For our purposes, the template $\mathcal{H}$ is prepared to match the point-like-source shape used before, and $D$ denotes in this context the astrophysical signal embedded in the noisy data sets described in previous sections.

Although the MF can recover compact-astronomical sources above a noise threshold, it is well known in general that the MF has a few limitations \citep{Allen2012}: it is not robust against non-white noise artifacts, and as the signal is not precisely known, the filter output SNR is strongly dependent on the template. Thus, MF is expected to under-perform our trained network, especially with highly structured noise, and yield more false positives from low-SNR raw data.

\section{Results}
\label{sec:results}

\begin{figure}
\includegraphics[width=\linewidth]{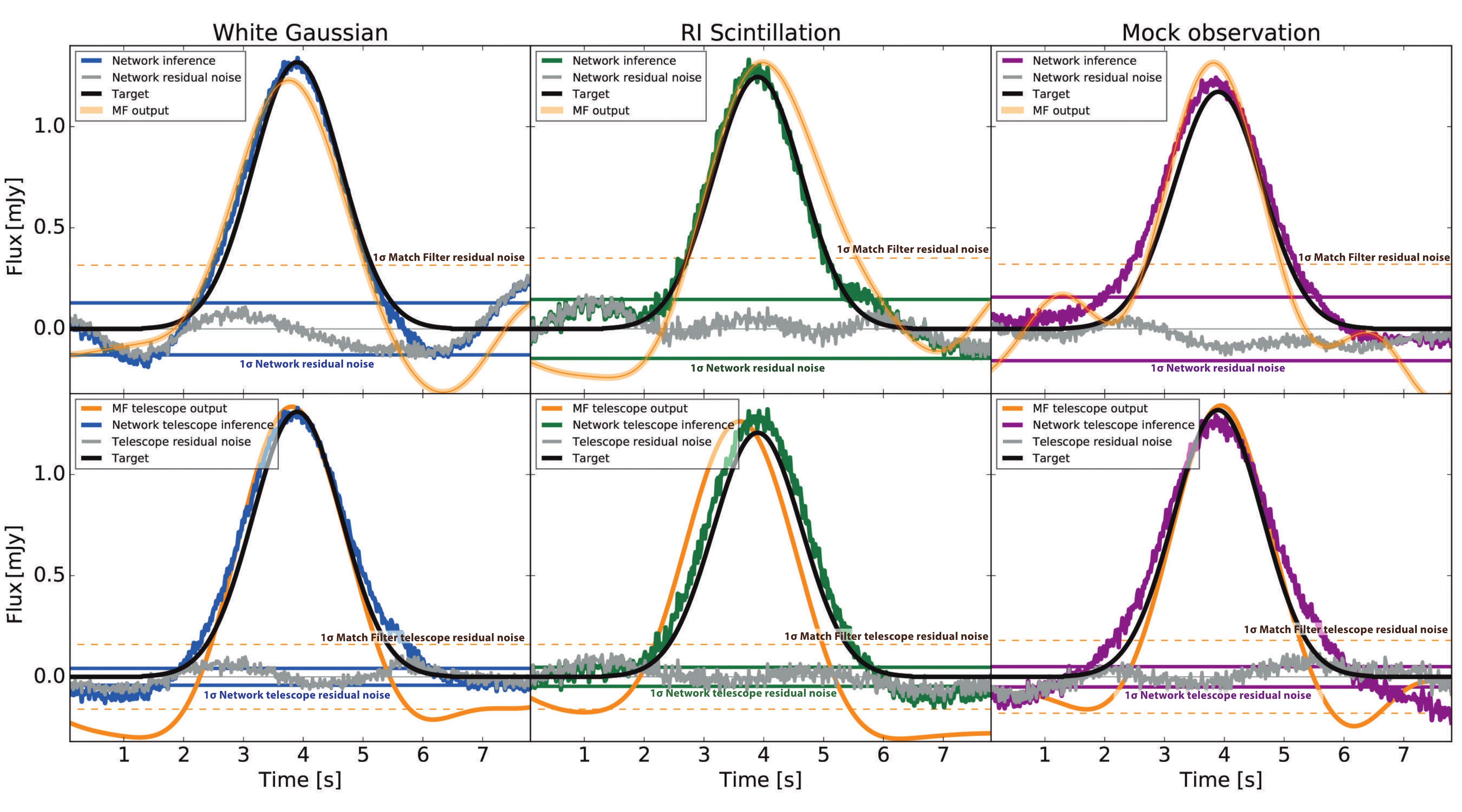}
\caption{
Upper panel: \textit{Single-shot} inferences of a $\sim$1.3 mJy source made by our Network and MF.
While in Fig. \ref{fig:inferences}, gray lines denote raw atmospheric noise, here, gray lines represent atmospheric noise residuals (denoted by $\hat{A}_i$ in the text) after our network-based filtering. (MF noise residuals are not shown.)
The horizontal color-solid lines represent the 1$\sigma$ network-residual noise level in the complete batch, while the orange-dashed lines are the size of the MF-residual noise. The network inferences (denoted as $\hat{s}_i$) are the colored noisy lines near the target ($s_i$), and the MF inferences are the orange-solid lines.
Lower panel: \textit{Telescope measurements} of the same sources after averaging 50 inferences of the same signal embedded in different noise realizations.
The horizontal-colored lines represent corresponding 1$\sigma$ residual noise levels in the complete batch. The telescope noise reduction from single network inferences accounts to roughly 66\% in these examples. This SNR is regarded as an approximation of the one that could be attained by a full-array integration in map-domain (see also Fig. \ref{fig:snr_snr}).
}
\label{fig:inferences_zoom}
\end{figure}

We are now interested in performing a statistical characterization of the network's inferences through incremental complexity levels. Including White Gaussian and IR Scintillation noise to this characterization is instructive to contextualize the results obtained with Mock Observations.
Fig. \ref{fig:inferences} shows three examples of network inferences with the three types of noise; 
altogether, they exemplify the network's inference of a faint source (feeding-SNR=0.8) embedded in incremental levels of structured noise.
Another set of time-streams with higher feeding-SNR is shown in the upper panel of Fig. \ref{fig:inferences_zoom}, focusing on the detection windows.

First, we need to define a criterion for a \textit{window-detection} (in a single-shot).
We see that for any given realization of atmospheric noise $A_i$ and an astrophysical point-like source $s_i$, the network inference $\hat{s}_i$ is a noisy representation of the true signal $s_i$.
Notice that low-frequency modes could dominate the residual atmospheric noise.
To estimate this residual noise, we feed the network with $A_i$ (that is equivalent to zero signal) and obtain an inference dubbed as $\hat{A}_i$, which is an approximation to the residual noise pattern contained in $\hat{s}_i$.
Henceforth, we denote the level of residual noise as std($\hat{A}_i$),
and look for peaks in $\hat{s}_i$ above 3 $\times$ std($\hat{A}_i$), labelling them as detection candidates.
Since we know beforehand our simulated signal, we can classify each detection as a \textit{true-positive} or \textit{false-positive}.
We also verify that the inference matches the target temporal location, allowing slight temporal displacements within a tolerance of $\sim$ 0.17 s, which is near to one-fourth of the source width.

\begin{figure}
    \centering
    \includegraphics[width=0.5\textwidth]{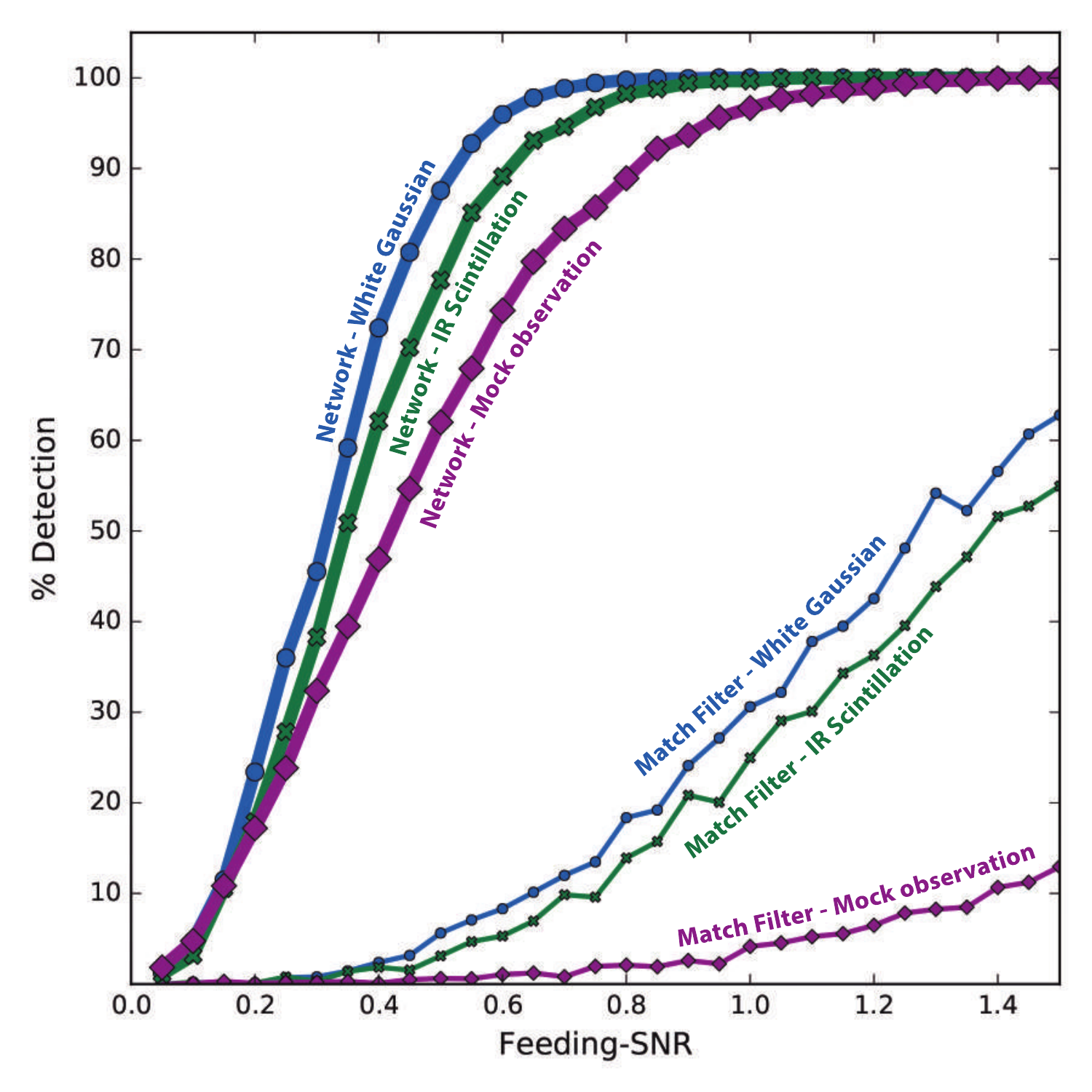}
    \caption{Source detection rate as a function of feeding-SNR.
    The curves quantify the fraction of the total number of simulated sources that are successfully cleaned and detected above a $3\sigma$ noise level after a single-shot Network or MF inference.
    Our neural network recovers a more significant fraction of sources than the MF. 
    For example, the MF completeness for a feeding-SNR$=$1 is $\simeq 35\%, 34\%$, and $4.4\%$ for White Gaussian, IR Scintillation, and Mock Observations, respectively. In contrast, our neural network produces a source detection above $95\%$ for the same feeding-SNR.}
    \label{fig:mf_comp}
\end{figure}

Using this criterion, we can count the number of sources successfully detected after a single inference over each structured noise type.
We generate a large number of new White Gaussian, IR Scintillation, and Mock Observation noisy datasets, each made of 150,000 realizations, created with feeding-SNR=0.1, 0.2, $\dots$, 3.5.
In Fig. \ref{fig:mf_comp}, we see the recovery rate as a function of feeding-SNR; it is higher for the less structured noise, reaching 95\% slightly above a feeding-SNR of 0.6 for White Gaussian, 0.8 for IR Scintillation, and 1 for Mock Observations. Besides, we find false-positive counts below 10\% at feeding-SNR $\lesssim$ 0.3.

To assert a comparison to a classical method, we test our MF implementation (section \ref{sec:mf}) over another set of 150,000 random realizations per each type of noise and compute the source detection rate in Fig. \ref{fig:mf_comp}. As it might have been anticipated, we see that the MF technique cannot recover as many astronomical sources as our trained network, especially at low SNR.
For example, with a feeding-SNR equal to one, the MF detection rates are
$\simeq$ 35\% for White Gaussian noise, 
34\% for IR Scintillation, and
4.4\% for Mock Observations.
These figures are in sharp contrast to our neural network results, generating a detection rate above $90\%$ for the same feeding-SNR (see Fig. \ref{fig:mf_comp}).
Besides, with MF, we find $\sim20\%$ false positives at feeding-SNR$\lesssim$1.

Even when an astrophysical source is recovered in a single-shot inference (as Fig. \ref{fig:mf_comp} shows), it does not necessarily imply an accurate measurement. 
We know that atmospheric noise residuals harm single-shot temporal samplings, causing distorted flux measurements. 
However, given the random nature of atmospheric noise, residuals are expected to introduce random departures from the actual flux, which, on average, should cancel-out at every pixel location in map-domain.

Henceforth, we use the term \textit{telescope measurement} as the average of a large number of stacked inferences in time-domain.
This term is reminiscent of what happens in map-domain: where the final measurement of an astronomical source is mainly the average of many (suitably cleaned) temporal samples. 
Consequently, although we do not address the map-making problem in our current analysis, we can, nonetheless, assert the expected map-domain detection quality.

Fig. \ref{fig:inferences_zoom} (lower panel) shows the telescope measurements of a $\sim$1.3 mJy point-like source after 50 stacked inferences. 
If the noise residuals were unstructured, 
the noise reduction after stacking 50 network inferences would have been $\sim 1/\sqrt{50}$,
or $\sim$85\%. 
However, we record an improvement of roughly 66\%, 
indicating that the residual noise left by our network still has some degree of structure.
We observe that the peak position is accurately recovered, 
though the inferred point-like source is broadened compared to the original target,
clearly as an accumulated effect of small temporal shifts around the signal's peak.
Atmospheric effects well known in astronomy as \textit{blurring} or \textit{wandering} are similar to the observed accumulated beam broadening \citep{Tatarskii2016wave,Tatarskii1971,Wheelon2001}.
For comparison, MF inferences in Fig \ref{fig:inferences_zoom} are significantly less accurate;
even after being stacked to form telescope measurements,
the average profile loses information at the skirts of the flux distribution.
Moreover, notice that our network's $1\sigma$ residual noise is much smaller than the level of MF residual noise in Fig. \ref{fig:inferences_zoom}.

\begin{figure}
\centering
  \includegraphics[width=\linewidth]{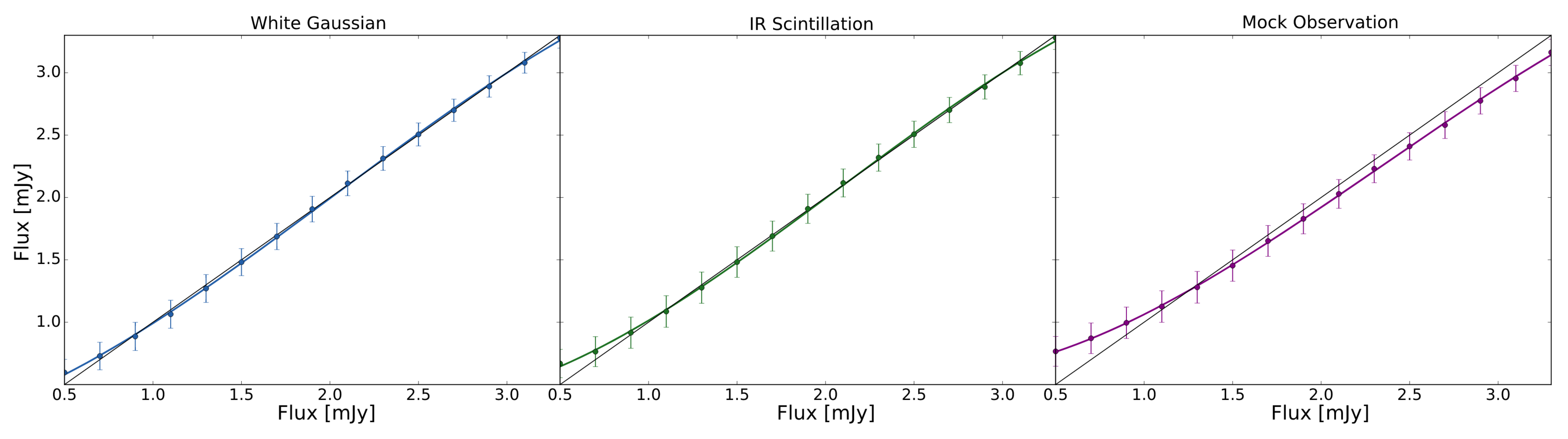}
  \caption{Flux calibration of network inferences. For each flux-bin, 2000 network inferences are stacked, and an average signal is statistically inferred. The error bars represent the uncertainty in the amplitude of the point-like source signal. The error-bars joining curve represent a polynomial fit that can be used to calibrate posterior network inferences.}
  \label{fig:calibration}
\end{figure}

Some sort of calibration is typically necessary for any cleaning methodology (see \textit{e.g.,} Refs. \citet{Downes2012,Rodriguez2018}). We can use telescope measurements not only to
characterize the network's inferences but even more important to design a flux calibration procedure, which will be critical by the time when real observational data from any new generation instrument will be available and processed by a scaled version of our current architecture.
We split the flux range into 14 bins, and generate 2,000 noise realizations for each flux bin.
We perform a Gaussian-curve fit to each averaged signal to measure the source flux and statistical error. From the 14 telescope measurements obtained, as shown in Fig. \ref{fig:calibration}, we see that they differ from the ideal one-to-one flux-correspondence line. Then, we perform a polynomial fit that we can use to calibrate subsequent network detection.

\begin{figure}
\includegraphics[width=\linewidth]{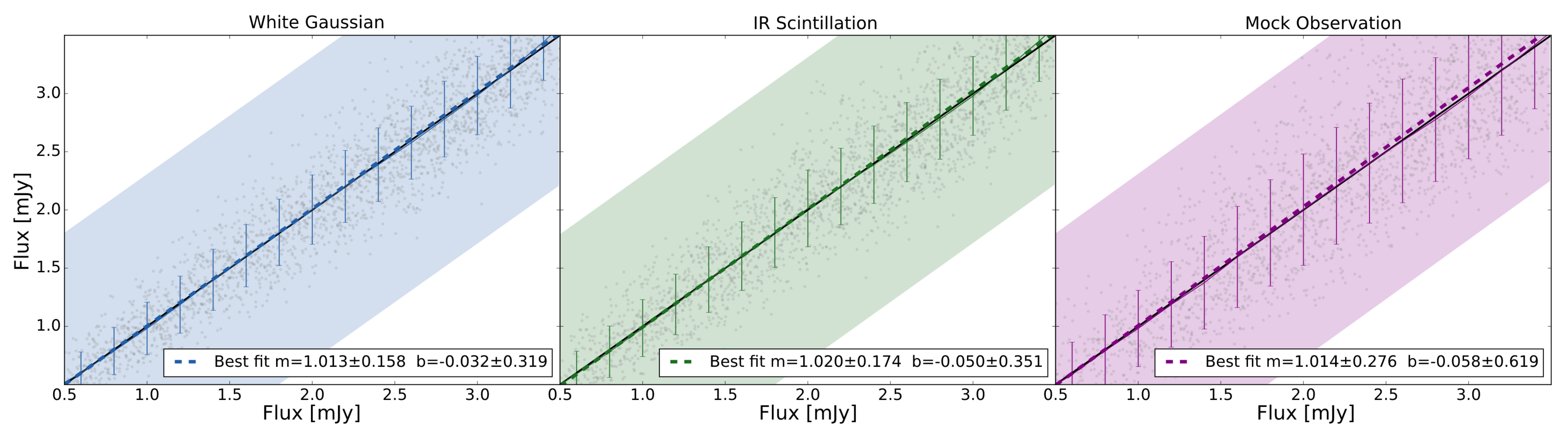}
\caption{Comparison of network-inferred source fluxes versus the real ones for the three types of simulated noise. Shaded regions represent 1$\sigma$ dispersion of the feeding noise.
Vertical error bars represent the inferred-fluxes 1$\sigma$ dispersion (per flux bin of 0.2 mJy) due to atmospheric residuals. The cloud of dots is a sub-representative sample of the total 150,000 inferences per each type of noise.
}
\label{fig:pointpoint}
\end{figure}

\begin{figure}
 \centering
  \includegraphics[width=0.5\linewidth]{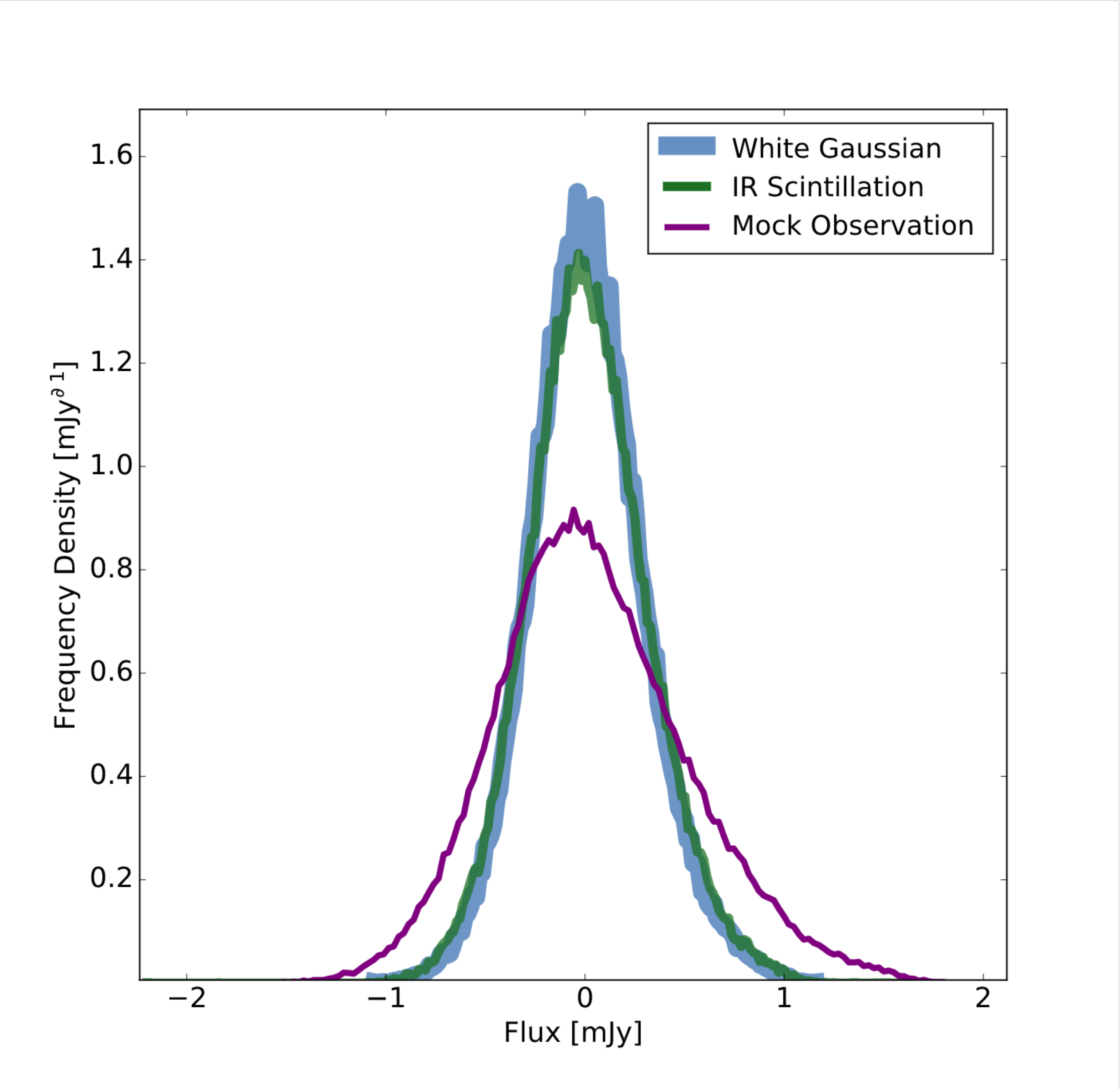}
   \caption{Distributions of random flux distortions on network inferences caused by atmospheric noise residuals, using the three types of structured noise.}
  \label{fig:density}
\end{figure}

To verify the calibration effectiveness, we utilize, once again, our 150,000 noisy data realizations, but correct the network flux measurements with the telescope measurements best-fit. The resulting cloud of calibrated flux-points appears in Fig. \ref{fig:pointpoint}. Vertical error bars denote the 1$\sigma$ dispersion amount. Shaded regions represent the typical 1$\sigma$ feeding-error bar that was used to generate these simulations.
For each type of noise, the dispersion-to-feeding-error ratio is
$\lesssim$ 0.27 for White Gaussian,
$\lesssim$ 0.29 for IR Scintillation, and
$\lesssim$ 0.44 for Mock Observations.
Performing a linear fit to the calibrated cloud of flux-points, we find a good agreement with the ideal one-to-one correspondence line, indicating a good calibration.
Subtracting the true flux and drawing the corresponding histograms on Fig. \ref{fig:density}, we can observe that the random distortion distributions tend to be symmetric, with a skewness-to-std ratio of
$\lesssim$ 0.70 for White Gaussian,
$\lesssim$ 0.63 for IR Scintillation, and
$\lesssim$ 0.70 for Mock Observations.
Thus, as we anticipated, we can assert that random-flux distortions will indeed cancel-out in the map-domain, provided that they have been correctly calibrated.
 
From Figs. \ref{fig:inferences_zoom}, \ref{fig:calibration}, and \ref{fig:pointpoint}, we see that Mock Observations display comparatively lower recovery rates, larger flux bias, and more scattered dispersion with respect to those of White Gaussian and IR Scintillation noise. We attribute this behavior to the structure degree contained in Mock Observations noise and the underlying instrumental systematic features that are likely present in these more realistic simulations.
Notwithstanding, in preliminary set-ups, we also identified that the training strategy could reduce both systematics. Indeed, to compensate for the higher complexity of low feeding-SNR samples, our TCL-SR training strategy weighs more on the latest reviewing stages, having lower flux sources. This scheme increases the recovery of weak sources (Fig. \ref{fig:mf_comp}), reduces the random distortions dispersion (Fig. \ref{fig:pointpoint}), and explains the over-estimations at low fluxes in Fig. \ref{fig:calibration} (which we correct through flux calibration).

\begin{figure}
\centering
\includegraphics[width=0.5\linewidth]{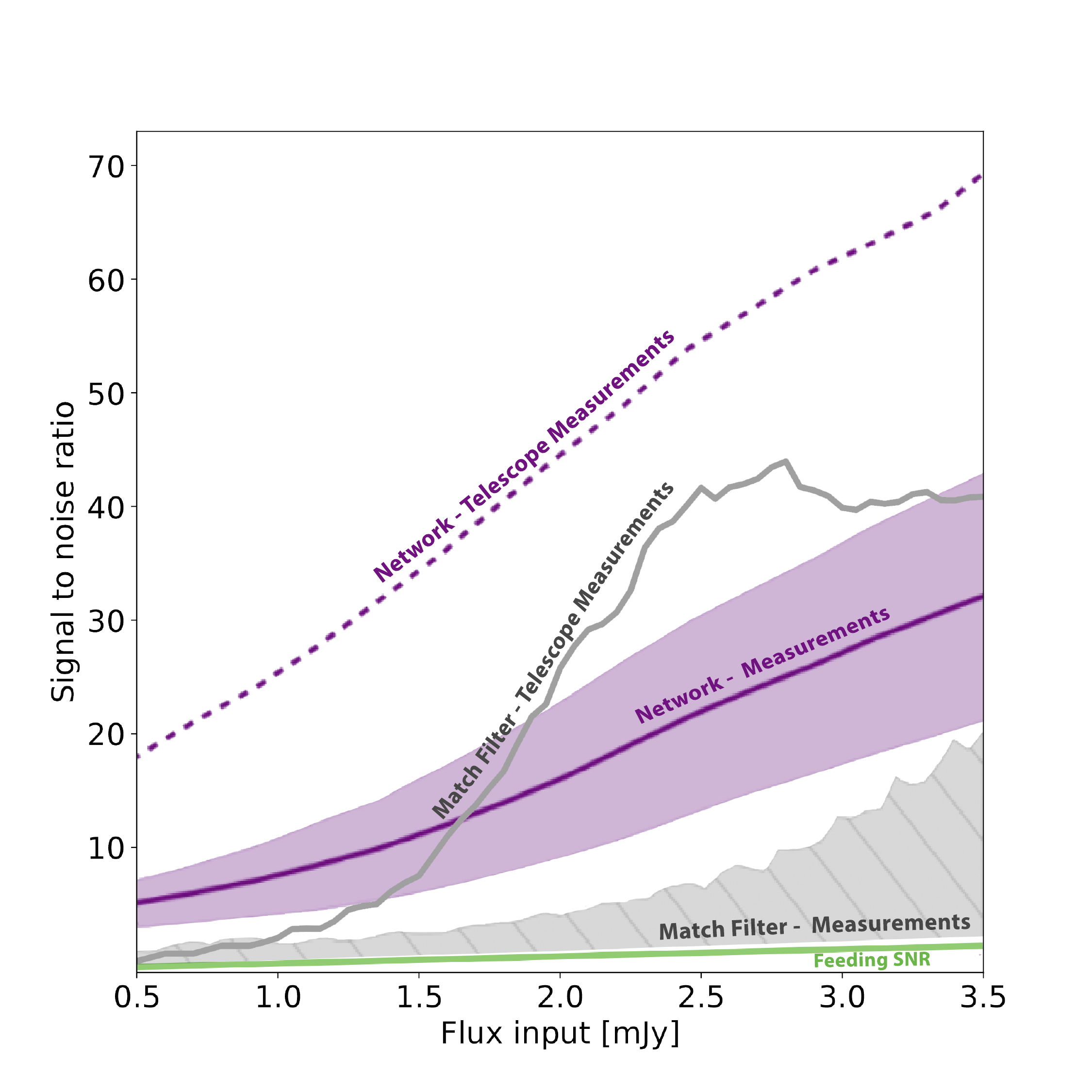}
  \caption{
  SNR enhancement from Mock Observation noisy series. The ensemble average upon 150,000 independent realizations is shown to exhibit the Network and MF detectability improvement. The curves represent the incrementally larger SNR per source flux in the case of raw data (feeding-SNR), Network and MF single-shot inferences, and Telescope measurements.}
  \label{fig:snr_snr}
\end{figure}

On the other hand, notice the significant noise-level drop in the telescope measurements in Fig.  \ref{fig:inferences_zoom}. This improvement is more clearly represented in Fig. \ref{fig:snr_snr}, where we show the SNR obtained from Mock Observations (only) as a function of flux. The single measurements' SNR is computed as the expectation value of the total 150,000 inferences (previously used for Fig. \ref{fig:mf_comp}). The telescope measurements' SNR is computed directly from their corresponding error bars (see Fig. \ref{fig:calibration}).
Both MF and our neural network inferences are represented in Fig. \ref{fig:snr_snr}.
For example, we see that a point-like source of 1 mJy, recorded in Mock Observations raw data with feeding-SNR=1.0, is expected to be measured by our neural network with an SNR between $\sim$5--13.
Likewise, after enough samples, the same source is expected to be measured by the telescope with an SNR of $\sim$27.
In contrast, a single MF inference of a 1 mJy source is expected with a SNR below $\sim$2.5, while a MF-telescope measurement at the same flux would only slightly improve the SNR. 
MF-telescope measurements show a maximum SNR at fluxes above $\sim$2.5 mJy. This plateau is explained by the information loss at the skirts of the MF inference profile (as shown in Fig. \ref{fig:inferences_zoom}), which is intensified for brighter sources.
Overall, our trained network reports a significant SNR enhancement of astrophysical sources with respect to raw signals, and superior to our tested MF technique.

As a side note, it is worth mentioning that our network obtained well-behaved inferences on temporal series generated from a spectral density absent in the training datasets. Given the variability of climate conditions along different observation nights, this capability to infer from diverse noise structures is an advantageous transfer learning capability to be further explored.

We should finally specify that our model was built in \textsc{tensorflow} \citep{abadi2016tensorflow} and trained in a tower-data configuration. The acceleration hardware consists of four Volta-100. The total training procedure required nearly 144 hours of GPU-time. The final frozen model takes 520 Mb in memory.

\section{Conclusions}
\label{sec:conclusion}

Structured noise in time-domain poses a significant challenge for communications and astronomy, especially under low-SNR conditions. 
Here, we have shown that a deep-learning approach with a representative training dataset could significantly advance the treatment of structured noise.
This contribution is of great interest for ground-based infrared and mm-wave astronomy, in the advent of a new technological generation of continuum cameras and large telescopes, because the turbulent Earth's atmosphere introduces highly structured noise that can be removed only through advanced data processing techniques and physical modeling.

We propose a bidirectional recurrent neural network model with the ability to abstract long- and short-term non-linear features from sequential data.
Our architecture consists of bidirectional LSTM cells that assemble a many-to-many input/output configuration. We utilize sliding windows of atmospheric noise with an embedded astrophysical signal to feed and train the network to retrieve a clean version of the signal. This configuration allows us to capture some of the larger atmospheric fluctuations while optimizing memory space and processing capabilities.

Because neural networks need to learn from large amounts of representative data, we develop and present here a complete model for the IR Scintillation spectral density, taking into account the Kolmogorov-Obukhov theory, the dry-air turbulent fluctuations (real part of the refractive index), and the anomalous dispersion caused by water-molecule emission lines (imaginary part of the refractive index).
Furthermore, we generate synthetic observations to train the network with a realistic type of noise, containing intrinsic structures induced possibly by instrumental effects.
We have tried four training strategies and adopt our TCL-SR strategy, a useful combination of curriculum and transfer learning with intermediate reviewing blocks.

We test our trained network with several complexity levels, reporting more than 95\% of source recovery (see Fig. \ref{fig:mf_comp}) at feeding-SNR $\gtrsim$ 0.6, 0.8, and 1, in the cases of White Gaussian, IR Scintillation, and Mock Observational noise, respectively.
We characterize random flux distortions caused by atmospheric noise residuals, helping us design an efficient TCL-SR training strategy, and calibrate inferences against flux bias. After an effective calibration, random flux distortions cancel out on average, leading to unbiased statistical measurements. 
It means that atmospheric noise residuals shall dilute in map-domain, provided a large enough number of temporal samples are stacked at every pixel.
Furthermore, our network reports a significant SNR enhancement of astronomical signals (see Fig. \ref{fig:snr_snr}). We notice that the network (single-shot) measurements improve the SNR from $\sim$5 to 44 times, on average, compared to raw data. Meanwhile, the telescope (statistical) measurements improve the SNR $\sim$19--71 times compared to raw data.

By comparison, our tests with the simple MF technique deliver a poor detection rate on single inferences and look nonrobust to handle different non-white Gaussian noise complexities \citep[more sophisticated versions like the adaptive MF may yield better results, though;][]{Robey92_AMF}. Thus, we assert that our neural network presents the potential of a much more efficient approach to converge to accurate astrophysical measurements than more traditional filtering methods.

We conclude that our deep-learning methodology reports a satisfactory cleaning performance over structured-noisy temporal series and considerably high astronomical-signal enhancement.
Additionally, our training strategy shows some evidence of effectiveness over diverse degrees of simulated noise structure (using spectral densities not contained in the training dataset).

It is worthy of stressing that although we have focused on point-like sources, they represent a wide variety of applications for extragalactic astronomy, ranging from telescope-pointing calibrations to the observation of far and faint sub-mm galaxies.
Regarding new generation instruments, both atmospheric and point-like source simulations are fairly representative realizations of the expected data. Nevertheless, more specialized simulations capturing each instrument's particular features may be generated from their observational data.

Our present study may be extended in many ways.
For instance, growing the LSTM's long-term memory could make it possible to abstract longer correlations to handle higher structure degrees or recover fainter sources.
Mechanisms such as Attention Models \citep{vaswani2017attention} or Compressive Transformers \citep{rae2019compressive} could provide interesting improvements as well.
Multi-feature networks may be another route for further studies, as it can be useful to train the network with environmental data taken at the observation Site.
For example, it would be useful to perform independent measurements of the temperature and specific-humidity structure parameters along the troposphere, which are well known to correlate with the recorded structured noise.
In this direction, our physical IR Scintillation model may be beneficial for further investigations attempting to disentangle the scintillation noise from a propagating signal through the Earth's atmosphere.
In turn, deep-learning research applied to atmospheric sciences could provide a parametric characterization of environmental variables from rough data, whose understanding is of great interest, not only for astronomy but also for a wide variety of fields where turbulence is involved. \\

\acknowledgments 
The authors thankfully acknowledge the computer resources provided by
the Laboratorio Nacional de Supercómputo del Sureste de México, 
CONACYT network of national laboratories.
This project was possible owing to partial support from CONACYT research grants 237004, 490769,
F.C.~2016/1848, and FORDECYT-297324. 
Finally, we would like to thank an anonymous referee for a critical review that improved significantly our paper.

\appendix

\section{Propagation of millimeter-waves through the turbulent atmosphere.}
\label{sec:propagation}

In this Appendix, we review atmospheric turbulence physics responsible for imprinting highly structured noise in the recorded signals. The Earth's atmosphere is a very complex medium; even during calm nights ideal for astronomical observations, turbulent fluctuations of wind velocity, air temperature, pressure, and specific humidity dominate the atmosphere dynamics. Although these thermodynamic variables are studied as climate phenomena, their small perturbations are entirely chaotic and make any deterministic description impossible.

Fortunately, the Kolmogorov's theory \citep{Pope2000,Nieuwstadt2016turbulence,Vallis2017atmospheric} provides a successful description of turbulent phenomena by distinguishing from three physical scales: \textit{i) the outer scale} $L_0$, characterized by large fluctuations (eddies) that are responsible for injecting mechanical energy into the system through progressive decays to smaller and smaller eddies. This energy cascade proceeds without (significant) loss down to \textit{ii) the inner scale} $l_0$, where, due to the air viscosity and molecular diffusion, all the energy is finally dissipated into heat.
\textit{iii) The universal inertial regime} is an intermediate range of longitudes $l_0 \ll l \ll L_0$ where dissipation is negligible and turbulence dynamics is locally homogeneous and isotropic. 
For the Earth's atmosphere, typically, $L_0\sim5-10$ km and $l_0\sim1$ mm.
Besides, at mm wavelengths, light interacts mainly along its path through the troposphere, which has a height of $H\sim\text{10 km}$. For a receiver pointing to a zenith angle $\theta$, the Fresnel length \citep{Wheelon2001,Wheelon2003} (that is comparable to the size of eddies causing light diffraction) is at most $\sqrt{\lambda H \sec\theta}$.
Since for typical observation angles, it is true that $l_0 \ll \sqrt{\lambda H \sec\theta} \ll L_0 $, the Kolmogorov's universal inertial regime provides a proper prescription of the atmospheric turbulence affecting mm astronomy.

\subsection{Wave equations and the turbulence description.}

The theory of wave propagation through turbulent media was developed in Tatarskii's (1961) seminal treatise \citep{Tatarskii2016wave}, followed by his own review \cite{Tatarskii1971}. Subsequent extensions have been summarized in more recent reviews \citep{Beland1993,Wheelon2001,Wheelon2003,Andrews2005}.
However, the fluctuations in the refractive index's imaginary part ---which account for absorption and re-emission (\textit{anomalous dispersion}) of light--- are often overlooked. Given that the former is a dominant effect at infrared and mm wavelengths, below, we shall review the equations governing the propagation of infrared light through a turbulent atmosphere.

An electromagnetic wave propagates through the atmosphere
obeying the wave equation,
\begin{eqnarray}
    \nabla^2 E(\bm{r}) + k^2 n^2(\bm{r}) E(\bm{r}) = 0,
    \label{eq:wave}
\end{eqnarray}
where $E$ is a component of the electric field, $k=|\mathbf{k}|$ is the wave-number amplitude and $n(\bf{r})$ is the atmosphere's refractive index.
The random nature of $n(\bm{r})$ prevents us to obtain any exact solution to the wave equation.
Nevertheless, approximate descriptions are still possible, like the Rytov approximation, which uses a transformation $E(\bm{r}) = E_0(\bm{r}) \, e^{\psi(\bm{r})}$ in order to perform a series expansion in terms of the surrogate function $\psi = \psi_0+\psi_1+\psi_2+\dots$,%
where the field $E_0$ obeys the vacuum wave equation.

We can describe the random behaviour of the refractive index in terms of small variations,
\begin{eqnarray}
    n(\bm{r}) = 1+n_1(\bm{r}),
\end{eqnarray}
where we have defined $\left\langle n(\bm{r})\right\rangle=1$.
In nature, $n_1(\bm{r}) = n_R(\bm{r}) - i n_I(\bm{r})$ is a complex passive scalar field whose real part is responsible for variations of the propagating wave due to refraction, and the imaginary part is responsible for absorption.
For visible light $n_I$ is usually negligible, meanwhile at infrared and mm wavelengths, the atmospheric molecules cause resonant absorption and re-emission of light; thus, anomalous dispersion is non-negligible.
Considering up to linear terms ($n^2\approx 1+2n_1$), we can see that the refractive index conveniently appears as the source in a first-order wave equation,
\begin{eqnarray}
    \nabla^2\psi_1(\bm{r}) + 2\nabla\psi_0(\bm{r})\cdot \nabla\psi_1(\bm{r}) = -2k^2 n_1(\bm{r}).
\end{eqnarray}
With an additional transformation $\psi_1(\bm{r})=Q(\bm{r})e^{-\psi_0(\bm{r})}$, one recovers the familiar Helmholtz equation, 
\begin{eqnarray}
    \nabla^2 Q + k^2 Q = -2k^2\,n_1(\bm{r})\; e^{\psi_0(\bm{r})},
\end{eqnarray}
whose solution is well known in terms of the Green function
$G(\bm{R},\bm{r}) = e^{i\bm{k}\cdot(\bm{R}-\bm{r})}/\left(4\pi |\bm{R}-\bm{r}|\right)$,
\begin{eqnarray}
  \psi_1(\bm{r}_o) = -2k^2\int d^3r\; G(\bm{R},\bm{r}) \frac{E_0(\bm{r})}{E_0(\bm{r}_o)}n_1(\bm{r}),
  \label{eq:rytov}
\end{eqnarray}
where $\bm{r}_o$ and $\bm{r}$ represent the observer and scattering positions respectively.
The first-order Rytov solution is readily found from
$E_1(\bm{r}_o) = E_0(\bm{r}_o)\; e^{\psi_1(\bm{r}_o)}$.
Implicitly, the former equations contain the expressions for the amplitude $A$ and phase $\varphi$ of the random-field fluctuation,
\begin{eqnarray}
    E_1(\mathbf{r}_o) = A(\bm{r}_o) \, \exp\left[i(\phi_0+\varphi(\mathbf{r}_o)\right],
\end{eqnarray}
where $\phi_0$ is the mean-field value of the phase fluctuations.
Yet, instead of the amplitude itself, it is commonly more useful to work with the log-amplitude $\chi\equiv\log(A/E_0)$.
Consequently, the astronomical signal fluctuations propagating through the atmosphere are described by the complex scattering integral
\begin{eqnarray}
    \chi(\mathbf{r}_o) + i\varphi(\bm{r}_o) = -2k^2 \int d^3r \;
     G(\bm{r}_o,\bm{r}) \frac{E_0(\bm{r})}{E_0(\bm{r}_o)}
    n_1(\bm{r}) .
\end{eqnarray}
Since the observed light intensity is the field square amplitude, $I/I_0 = |E_1|^2/|E_0|^2 = e^{2\chi}$, random variations of $\chi$ are a proxy to quantify the noise induced by the turbulent atmosphere; an effect also known as \textit{scintillation}.
Notice that as $\left\langle n_1 \right\rangle = 0$,
we also have $\left\langle\chi\right\rangle=0$ by definition.
However, the second-order statistical moments are non-negligible;
indeed, the scintillation noise variance can be computed from
\begin{eqnarray}
 \sigma^2_I=\frac{\left\langle [I-\left\langle I\right\rangle ]^2\right\rangle}{\left\langle I\right\rangle^2} 
 \approx 4\left\langle\chi^2\right\rangle ,
\end{eqnarray}
where,
\begin{eqnarray}
  \left\langle\chi^2\right\rangle &=& 4k^4\int d^3r\int d^3r'
  \left\langle
  \operatorname{Re}\left\lbrace
  G(\bm{r}_o,\bm{r})\frac{E_0(\bm{r})}{E_0(\bm{r}_o)}
  n_1(\bm{r})    \right\rbrace
  \operatorname{Re}\left\lbrace
  G(\bm{r}_o,\bm{r}')
  \frac{E_0(\bm{r}')}{E_0(\bm{r}_o)} 
  n_1(\bm{r}')  \right\rbrace
  \right\rangle .
   \label{eq:flucts}
\end{eqnarray}
To compute this double integral we need some prior knowledge of
the complex refractive index auto-covariance function
$\left\langle n_1(\bm{r})\,n_1(\bm{r}')\right\rangle$.
For cell sizes $l_0 \ll |\bm{r}-\bm{r}'| \ll L_0$, well inside the universal inertial regime,
we know that turbulence is homogeneous, thus,
the auto-covariance can only depend on the relative position $\bm{r}-\bm{r}'$
but not on the individual vectors $\bm{r}$ and $\bm{r}'$.
Furthermore, each component of the auto-covariance is associated to its Fourier power spectrum by,
\begin{eqnarray}
 \left\langle n_R(\bm{r}) \, n_R(\bm{r}') \right\rangle &=& \int d^3\kappa \;
 e^{i\bm{\kappa}\cdot (\bm{r}-\bm{r}')} \; \Phi_R(\bm{\kappa}), \nonumber \\
 \left\langle n_I(\bm{r}) \, n_I(\bm{r}') \right\rangle &=& \int d^3\kappa \;
 e^{i\bm{\kappa}\cdot (\bm{r}-\bm{r}')} \; \Phi_I(\bm{\kappa}), 
 \label{eq:ncov} \\
 \left\langle n_R(\bm{r}) \, n_I(\bm{r}') \right\rangle &=& \int d^3\kappa \;
 e^{i\bm{\kappa}\cdot (\bm{r}-\bm{r}')} \; \Phi_{IR}(\bm{\kappa}),  \nonumber
\end{eqnarray}
where $\bm{\kappa}$ is the eddy-wave-vector.
Moreover, provided the turbulence is isotropic inside the universal inertial regime, we know that $\Phi_n$ should not depend on any specific direction but just on the magnitude of the wave-vector $\kappa=2\pi/l$.
According to a simple dimensional analysis, Kolmogorov and Obukhov showed that
in this approximation $\Phi_n$ is given by the Kolmogorov-Obukhov (KO) spectrum,
$\Phi_n(\kappa) = 0.033 \, C^2_n \, \kappa^{-11/3}$
\citep{Kolmogorov1941,Obhukov1941},
where the structure constant $C^2_n$ must be determined experimentally.
For cell sizes barely comparable to the outer scale,
$|\bm{r}-\bm{r}'| \lesssim L_0$, the KO spectrum is still valid locally,
but $C_n^2$ needs to be corrected as a function of position.
For the Earth's atmosphere, the structure parameter depends
mainly on the height above the observer
$C^2_n(\bm{r}) \rightarrow C^2_n(z)$.
More generally, the turbulence spectrum can be characterized  by
\begin{eqnarray}
 \Phi_n(\bm{\kappa};z) = C^2_n(z) \; \Phi_0(\kappa), 
    \label{eq:KOspec}
\end{eqnarray}
where from here on the subscript $n$ denotes $R,I,$  or $IR$.
$\Phi_0(\kappa)$ could be written as the original KO spectrum $0.033\,\kappa^{-11/3}$;
but, since both the outer and dissipation scales suppress power
at the largest and smallest scales, respectively,
several empirical models have been developed in order to account for
deviations from the inertial regime;
\textit{e.g.}, the Tatarskii's (1961) spectrum \citep{Tatarskii2016wave},
the von Kármán's (1948) \cite{vonKarman1948}, and yet
a more versatile version (that contains the former two)
was proposed by Pope (2000) \cite{Pope2000},
\begin{eqnarray}
  \Phi_0(\kappa) = 0.033\,\kappa^{-11/3}f_L(\kappa L_0)\,f_l(\kappa l_0),   &
  \hspace{0.4cm} f_L(\kappa L_0) = \left(\frac{\kappa L_0}{(\kappa L_0)^2+c_L^2}\right)^{5/3+p_0}, &
  \hspace{0.4cm} f_l(\kappa l_0) = e^{-\beta_0\, \kappa l_0},
  \label{eq:Pope}
\end{eqnarray}
with $L_0$=5-10 km, $c_L=2.6$, and $p_0=4$ as some typical values.
Here we won't consider dissipation scale effects, so that we choose $\beta_0=0$.

A careful analysis shows that $C^2_R$ depends mainly on the fluctuations of temperature and humidity, while $C^2_I$ depends mainly on the humidity fluctuations \citep{Hill1980refractive}. Yet, it is not possible to determine any of the structure parameters from first principles and they have to be measured directly (see \textit{e.g.}, \citep{Osborn2016C2n,Qian2018AliObs,Avila2019SanPedro,Avila2021}).
For many years, this kind of experiments (in different locations and altitudes) have generated empirical profiles for $C_n^2(z)$; one of the most popular is the so-called Hufnagel-Valley 5/7 (HV5/7) model \citep{Beland1993},
\begin{eqnarray}
  C^2_n(z) = 1.7\times10^{-14}e^{-z/100}
  + 2.7\times10^{-16}e^{-z/1500}
  + 8.2\times10^{-26} \upsilon_{\rm rms}^2 \left(\frac{z}{1000}\right) e^{-z/1000} ,
  \label{eq:HV57}
\end{eqnarray}

where $\upsilon_{\rm rms}$ is the transverse r.m.s. wind speed.
Ideally, one needs especial measurements of the structure parameter
but these measurements are not always available.
Nonetheless, we can make use of an empirical profile like the HV5/7 model to make some approximations.
By substituting equations (\ref{eq:ncov}), (\ref{eq:KOspec}), (\ref{eq:Pope}), and (\ref{eq:HV57}) into (\ref{eq:flucts}), we can estimate the second-order statistical moments of the astronomical signal passing through the turbulent atmosphere.

\subsection{Scintillation variance.}
For simplicity, we may initially think of a point-receiver and take the origin at the receiver position $\bm{r}_o=0$. Consider the incident astronomical signal as a plane-wave propagating from the zenith $E_0(\bm{r})=\mathcal{E}_0\,e^{-ikz}$.
We can also use the approximation of a small scattering angle, which is valid under conditions of weak-scattering, then in cylindrical coordinates, $\sqrt{\rho^2+z^2}\approx z+\rho^2/2z$. The argument of the scattering integral is, in this case,
\begin{eqnarray}
  G(0,\bm{r})\frac{E_0(\bm{r})}{E_0(0)} n_1(\bm{r})  =
  \frac{1}{4\pi z}\left[
  n_R(\bm{r})\cos\left(\frac{k\rho^2}{2z}\right)
 +n_I(\bm{r})\sin\left(\frac{k\rho^2}{2z}\right)
 \right]
 +\frac{i}{4\pi z}\left[
  n_R(\bm{r})\sin\left(\frac{k\rho^2}{2z}\right)
 +n_I(\bm{r})\cos\left(\frac{k\rho^2}{2z}\right)
 \right] .
\end{eqnarray}
This can be substituted in Eq. (\ref{eq:flucts}) with the aid of (\ref{eq:ncov}) to compute the fluctuation auto-covariance, which is the sum of the real, imaginary, and the imaginary-real contributions,
\begin{eqnarray}
  \left\langle\chi^2\right\rangle = \left\langle\chi^2_R\right\rangle + \left\langle\chi^2_I\right\rangle + \left\langle\chi^2_{IR}\right\rangle .
\end{eqnarray}
Let us define the weighting functions
\begin{eqnarray}
  A_n(\bm{\kappa}) = \int d^3r\frac{C_n(z)}{z}\cos\left(\frac{k\rho^2}{2z}\right) 
  e^{i\bm{\kappa}\cdot\bm{r}}  ,  &&
  B_n(\bm{\kappa}) = \int d^3r\frac{C_n(z)}{z}\sin\left(\frac{k\rho^2}{2z}\right)
  e^{i\bm{\kappa}\cdot\bm{r}} ,
\end{eqnarray}
in order to express each component as
\begin{eqnarray}
  \left\langle \chi^2_R \right\rangle = \frac{k^4}{(2\pi)^2} \int d^3\kappa \;\Phi_0(\kappa) \; A_R(\bm{\kappa})\,A_R(-\bm{\kappa}), &&
  \left\langle \chi^2_I \right\rangle = \frac{k^4}{(2\pi)^2} \int d^3\kappa \;\Phi_0(\kappa) \; B_I(\bm{\kappa})\,B_I(-\bm{\kappa}),
  \label{eq:vars_expr}
\end{eqnarray}
\begin{eqnarray}
  \left\langle \chi^2_{IR} \right\rangle = 
  \frac{k^4}{(2\pi)^2} 
  \int d^3\kappa \;\Phi_0(\kappa) \; 
  \left[ A_R(\bm{\kappa})\,B_I(-\bm{\kappa}) + B_R(\bm{\kappa})\,A_I(-\bm{\kappa}) \right] \nonumber,
\end{eqnarray}
where $\Phi_0(\kappa)$ is given in ($\ref{eq:Pope}$).
The integrals over the angular coordinates are easily done under the assumption of local isotropy,
\begin{eqnarray}
  \left\langle\chi^2_R\right\rangle &=& (2\pi k)^2\sec\vartheta
  \int_{h_{\rm min}}^{h_{\rm max}}dz\,C^2_R(z)
  \int_0^\infty d\kappa\,\kappa\,\Phi_0(\kappa)\,
  \sin^2\left(\frac{\kappa^2 z\sec \vartheta}{2k}\right) , 
  \nonumber\\
  \left\langle\chi^2_I\right\rangle &=& (2\pi k)^2 \sec\vartheta
  \int_{h_{\rm min}}^{h_{\rm max}}dz\,C^2_I(z)
  \int_0^\infty d\kappa\,\kappa\,\Phi_0(\kappa)\,
  \cos^2\left(\frac{\kappa^2 z\sec \vartheta}{2k}\right) , 
  \label{eq:Chi_var} \\
  \left\langle\chi^2_{IR}\right\rangle &=& 2(2\pi k)^2 \sec\vartheta
  \int_{h_{\rm min}}^{h_{\rm max}}dz\,C^2_{IR}(z)
  \int_0^\infty d\kappa\,\kappa\,\Phi_0(\kappa)\,
  \sin\left(\frac{\kappa^2 z\sec \vartheta}{2k}\right)
  \cos\left(\frac{\kappa^2 z\sec \vartheta}{2k}\right)
  \nonumber .  
\end{eqnarray}
where we have made the line-of-sight change of variable $z\rightarrow s=z\sec\vartheta$, with $\vartheta$ the zenith angle, $h_{\rm min}$ is the altitude at which the observatory sits on, and $h_{\rm max}$ is the maximum relevant height, in our case, the end of the troposphere. 
Several investigations \citep{Hill1980refractive,Lee1969,Ludi2005,Solignac2012,Yuan2015} have shown that $C^2_R$ is typically $10^3$-$10^7$ times larger than $C^2_I$.
This might naively lead to the wrong assumption that the imaginary-part contributions would be negligible with respect to the purely-real part, but as we will see, this is not the case for infrared and mm-wavelengths.

\subsection{Covariance and aperture averaging.}
\begin{figure}
\includegraphics[width=\linewidth]{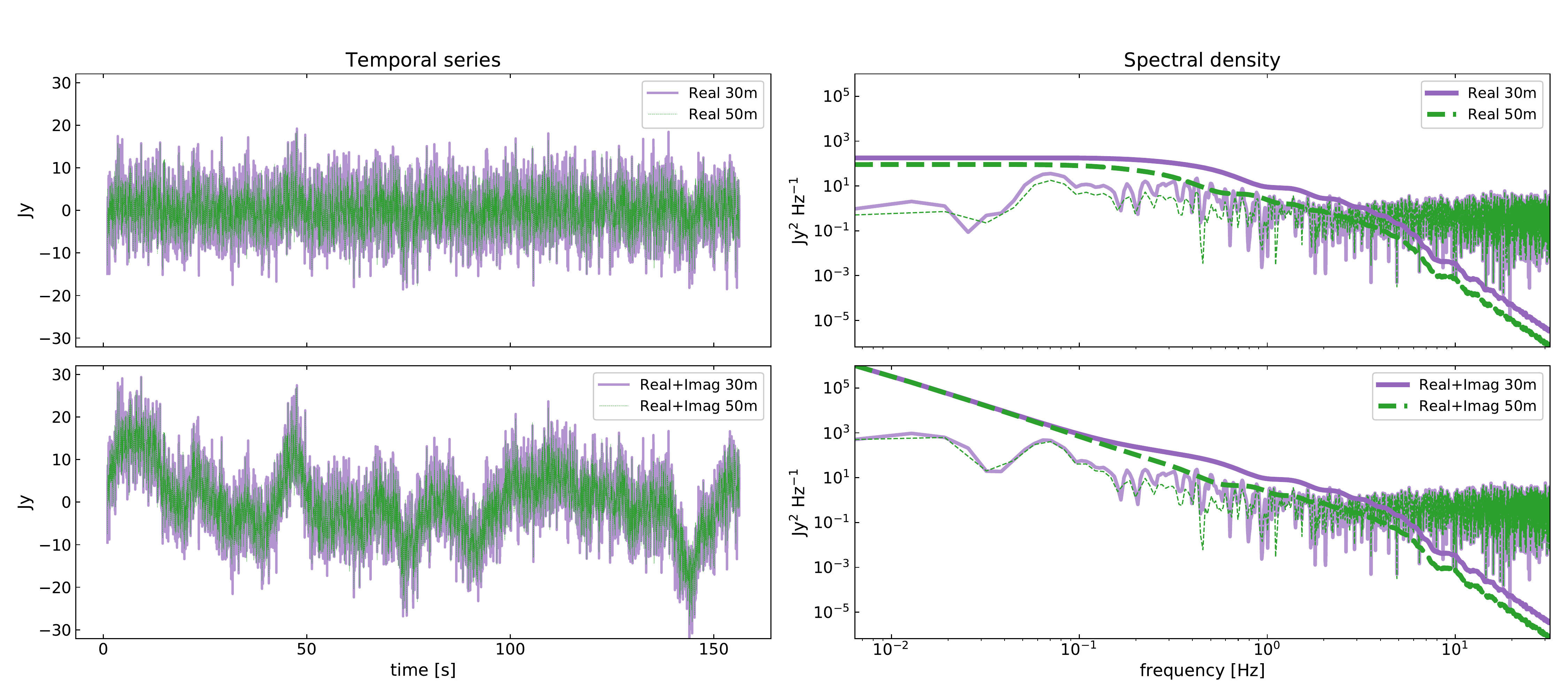}
\caption{Theoretical realizations of the scintillation noise (left)
         and its corresponding spectral density (right),
         caused by the atmospheric turbulence
         perturbing the electromagnetic waves
         propagating through the Earth's troposphere.
         The infrared and mm-waves are affected by the dry air
         ---or real refractive index---
         turbulent fluctuations (top panel);
         but more dominantly, they are affected by the water vapor 
         ---or imaginary refractive index---
         turbulent fluctuations (bottom).
         A circular telescope aperture with 30 and 50 m of diameter
         is considered to compute each noise simulation.
         Theoretical scintillation power spectra are represented by thick lines 
         (corresponding to $\mathcal{P}$ in Eq. \ref{eq:atm_sim}).
         Random-noise realizations are created according to Eq. (\ref{eq:atm_sim}) and
         represented by thin lines in the spectral densities and temporal series.
         }
\label{fig:theor_scintillation}
\end{figure}

Consider a pair of point-receivers separated by a (constant) baseline vector $\bm{\rho}_b$. The spatial covariance between two point-receivers is very similar to the previous computation in Eqs. (\ref{eq:vars_expr}),
\begin{eqnarray}
 \left\langle\chi_I(\bm{r}_o)\;\chi_I(\bm{r}_o+\bm{\rho}_b)\right\rangle =
 \frac{k^4}{(2\pi)^2}\int d^3\kappa
 \;\Phi_0(\kappa) \; e^{i\bm{\kappa}\cdot\bm{\rho}_b} \;
 B_I(\bm{\kappa})\,B_I(-\bm{\kappa}) . \nonumber 
\end{eqnarray}
The expressions for $R$ and $IR$ components are closely similar.
Now, the integration over the angular coordinates yields,
\begin{eqnarray}
\left\langle\chi_R(\bm{r}_o)\;\chi_R(\bm{r}_o+\bm{\rho}_b)\right\rangle &=&
   (2\pi k)^2\sec\vartheta
 \int_{h_{\rm min}}^{h_{\rm max}}dz\,C^2_R(z)
 \int_0^\infty d\kappa \,\kappa\,\Phi_0(\kappa) \, J_0(\kappa\rho_b) 
 \sin^2\left(\frac{\kappa^2 z\sec \vartheta}{2k}\right) , \nonumber \\
\left\langle\chi_I(\bm{r}_o)\;\chi_I(\bm{r}_o+\bm{\rho}_b)\right\rangle &=&
   (2\pi k)^2\sec\vartheta
 \int_{h_{\rm min}}^{h_{\rm max}}dz\,C^2_I(z)
 \int_0^\infty d\kappa \,\kappa\,\Phi_0(\kappa) \, J_0(\kappa\rho_b) 
 \cos^2\left(\frac{\kappa^2 z\sec \vartheta}{2k}\right) ,
 \label{eq:spat_cov} \\
 \left\langle\chi_R(\bm{r}_o)\;\chi_I(\bm{r}_o+\bm{\rho}_b)\right\rangle &=&
   2(2\pi k)^2\sec\vartheta
 \int_{h_{\rm min}}^{h_{\rm max}}dz\,C^2_{IR}(z)
 \int_0^\infty d\kappa \,\kappa\,\Phi_0(\kappa) \, J_0(\kappa\rho_b) 
 \sin\left(\frac{\kappa^2 z\sec \vartheta}{2k}\right)
 \cos\left(\frac{\kappa^2 z\sec \vartheta}{2k}\right), \nonumber
\end{eqnarray}
where $J_0$ is the zeroth-order spherical Bessel function.
Using (\ref{eq:spat_cov}), we want to correct the scintillation variance in (\ref{eq:Chi_var}) by considering the electromagnetic-waves collecting-average of a large telescope.
To do so, we can assume for simplicity a circular dish of radius $a_r$ (more realistic aperture corrections can be considered \citep{Osborn2015origins,Osborn2015dishcorrection}).
Then, we need to sum over all point-receiver spatial covariances located at pair-positions $\bm{\rho}_1$ and $\bm{\rho}_2$ inside the circular aperture, such that their (now variable) baseline distance
$\rho_b=|\bm{\rho}_1-\bm{\rho}_2|=\left[\rho_1^2+\rho_2^2-2\rho_1\rho_2\cos(\phi_1-\phi_2)\right]^{1/2}$
is $\rho_b \leq a_r$. Then, the aperture-average scintillation variance, defined by
\begin{eqnarray}
 \overline{ \left\langle\chi^2\right\rangle } &=& 
 \frac{1}{\pi^2a_r^2}\int_0^{a_r}d\rho_1\,\rho_1\int_0^{a_r}d\rho_2\,\rho_2
 \int_0^{2\pi}d\phi_1\int_0^{2\pi}d\phi_2 
 \left\langle\chi(\rho_1,\phi_1)\;\chi(\rho_2,\phi_2)\right\rangle , \nonumber 
\end{eqnarray}
can be computed for each component,
\begin{eqnarray}
 \overline{ \left\langle\chi^2_R\right\rangle } &=& 
    (2\pi k)^2\sec\vartheta
  \int_{h_{\rm min}}^{h_{\rm max}}dz\,C^2_R(z)
  \int_0^\infty d\kappa\,\kappa\,\Phi_0(\kappa)
  \sin^2\left(\frac{\kappa^2 z\sec \vartheta}{2k}\right)
  \left(\frac{2J_1(\kappa a_r)}{\kappa a_r}\right)^2 , \nonumber \\
   \overline{ \left\langle\chi^2_I\right\rangle } &=& 
    (2\pi k)^2\sec\vartheta
  \int_{h_{\rm min}}^{h_{\rm max}}dz\,C^2_I(z)
  \int_0^\infty d\kappa\,\kappa\,\Phi_0(\kappa)
  \cos^2\left(\frac{\kappa^2 z\sec \vartheta}{2k}\right)
  \left(\frac{2J_1(\kappa a_r)}{\kappa a_r}\right)^2 , 
  \label{eq:apaverage} \\
   \overline{ \left\langle\chi^2_{IR}\right\rangle } &=& 
    2(2\pi k)^2\sec\vartheta
  \int_{h_{\rm min}}^{h_{\rm max}}dz\,C^2_{IR}(z)
  \int_0^\infty d\kappa\,\kappa\,\Phi_0(\kappa)
  \sin\left(\frac{\kappa^2 z\sec \vartheta}{2k}\right)
  \cos\left(\frac{\kappa^2 z\sec \vartheta}{2k}\right)
  \left(\frac{2J_1(\kappa a_r)}{\kappa a_r}\right)^2 , \nonumber 
\end{eqnarray}
where $J_1$ is the first-order spherical Bessel function.
The overall effect of a large dish is to reduce the scintillation noise. 
But notice that the $\sin^2(x)$ function has significantly less overlap with $J_1^2(x)/x^2$ than $\cos^2(x)$. Consequently, the noise reduction is quite notable for the refractive-index real part; however, the scintillation noise due to the imaginary part remains almost unchanged irrespective of dish size (see Fig. \ref{fig:theor_scintillation}).

\subsection{Temporal power spectrum.} 
For the next step, we need to invoke the \textit{Taylor's frozen turbulence hypothesis} \citep{Taylor1938},  which postulates that for short enough periods of time and for slowly changing wind velocities, the turbulent atmosphere can be seen as a frozen screen moving in an horizontal plane along the line-of-sight.
In other words, the Taylor's hypothesis states an equivalence between spatial and temporal covariances, allowing us to make the replacement $\rho_b\rightarrow \tau\upsilon$ in equation (\ref{eq:spat_cov}), in terms of the wind velocity transverse component $\upsilon(z)$.
Thus, the temporal covariance can be written as,
\begin{eqnarray}
\overline{\left\langle\chi_R(t)\;\chi_R(t+\tau)\right\rangle} &=& 
   (2\pi k)^2\sec\vartheta
  \int_{h_{\rm min}}^{h_{\rm max}}dz\,C^2_R(z)
  \int_0^\infty d\kappa\,\kappa\,\Phi_n^{(0)}(\kappa)
  \,J_0[\kappa\tau\upsilon(z)]\,
  \sin^2\left(\frac{\kappa^2 z\sec \vartheta}{2k}\right)
  \left(\frac{2J_1(\kappa a_r)}{\kappa a_r}\right)^2 , 
  \label{eq:temp_cov}
\end{eqnarray}
with similar expressions for the $I$ and $IR$ components,
and where we already include the aperture-averaging factor.
The frequency power density, defined by
$\overline{\left\langle\chi^2\right\rangle}=\int d\omega \,P_\chi(\omega)$,
can be computed from the Fourier transformation of the temporal covariance,
\begin{eqnarray}
    P_\chi(\omega) = \int_{-\infty}^\infty d\tau\;e^{i\omega\tau}\;
    \overline{\left\langle\chi(t)\;\chi(t+\tau)\right\rangle} .
\end{eqnarray}
The full power spectrum is the sum of the complex refractive index contributions,
$P_\chi=P_R(\omega)+P_I(\omega)+P_{IR}(\omega)$.
The temporal integration can be done analytically under the condition
$\omega<\kappa\upsilon<\infty$, leading us to
\begin{eqnarray}
 P_R(\omega) &=& 
  2(2\pi k)^2\sec\vartheta
  \int_{h_{\rm min}}^{h_{\rm max}}dz\,C^2_R(z)
  \int_{\omega/\upsilon(z)}^\infty d\kappa\,
  \frac{\kappa\,\Phi_0(\kappa)}{\sqrt{(\kappa\upsilon)^2-\omega^2}}
  \sin^2\left(\frac{\kappa^2 z\sec \vartheta}{2k}\right)
  \left(\frac{2J_1(\kappa a_r)}{\kappa a_r}\right)^2 , \nonumber \\
   P_I(\omega) &=&
  2(2\pi k)^2\sec\vartheta
  \int_{h_{\rm min}}^{h_{\rm max}}dz\,C^2_I(z)
  \int_{\omega/\upsilon(z)}^\infty d\kappa\,
  \frac{\kappa\,\Phi_0(\kappa)}{\sqrt{(\kappa\upsilon)^2-\omega^2}}
  \cos^2\left(\frac{\kappa^2 z\sec \vartheta}{2k}\right)
  \left(\frac{2J_1(\kappa a_r)}{\kappa a_r}\right)^2 , \label{eq:app_specs} \\
   P_{IR}(\omega) &=&
  4(2\pi k)^2\sec\vartheta
  \int_{h_{\rm min}}^{h_{\rm max}}dz\,C^2_{IR}(z)
  \int_{\omega/\upsilon(z)}^\infty d\kappa\,
  \frac{\kappa\,\Phi_0(\kappa)}{\sqrt{(\kappa\upsilon)^2-\omega^2}}
  \sin\left(\frac{\kappa^2 z\sec \vartheta}{2k}\right)
  \cos\left(\frac{\kappa^2 z\sec \vartheta}{2k}\right)
  \left(\frac{2J_1(\kappa a_r)}{\kappa a_r}\right)^2 . \nonumber
\end{eqnarray}
These integrals can be easily evaluated numerically once some profiles for the structure parameters and wind speed have been specified. 
The Bufton's wind profile \citep{Andrews2005} is a simple empirical model suggesting a Gaussian shape for $\upsilon(z)$, with a stream peaking barely below the troposphere's height.
Notice that the structure parameter integration over $z$ contributes effectively as an overall numerical pre-factor; thus, we can use equation (\ref{eq:HV57}) as the altitude profile for the three structure parameters, and modulate parametrically their relative power.

From the theoretical scintillation spectral density in Eqs. (\ref{eq:app_specs}), we can generate many simulations of atmospheric temporal series using Eq. (\ref{eq:atm_sim}). Fig. \ref{fig:theor_scintillation} exemplifies these random realizations, with 30 and 50 m telescope aperture diameters, and observations at the zenith. The upper-left panel shows two dry-air scintillation (real refractive index) simulated time-stream. The corresponding theoretical spectral densities are represented by the smooth curves on the upper-right panel, which are convolved with a Gaussian-random series to simulate an observation.
The lower-left panel shows two wet-air scintillation (real+imaginary refractive index) simulations with $C^2_I=10^{-5}C^2_R$. The lower-right panel shows the corresponding theoretical and random spectral densities. From the theoretical spectral density, we observe a high-frequency power reduction by the 50 m aperture compared to 30 m; it means that a large telescope can notably reduce the dry-air scintillation noise. On the other hand, the low-frequency power remains equal for both 30 and 50 m; it means that a larger dish cannot defeat the wet-air scintillation noise.

In summary, Eqs. (\ref{eq:app_specs}) represent the general scintillation spectral density at infrared and mm-waves, derived from turbulence-physics and telescope aperture-averaging factors. The large-scale highly-structured noise prevailing in the observational temporal series is mainly caused by the amplitude of the imaginary-part structure parameter $C_I(z)$. This derivation is presented for an astronomical application for the first time and should be the preferred model to simulate atmospheric noise at infrared and mm-waves.


\bibliography{refs}

\begin{thebibliography}{}
\expandafter\ifx\csname natexlab\endcsname\relax\def\natexlab#1{#1}\fi
\providecommand{\url}[1]{\href{#1}{#1}}
\providecommand{\dodoi}[1]{doi:~\href{http://doi.org/#1}{\nolinkurl{#1}}}
\providecommand{\doeprint}[1]{\href{http://ascl.net/#1}{\nolinkurl{http://ascl.net/#1}}}
\providecommand{\doarXiv}[1]{\href{https://arxiv.org/abs/#1}{\nolinkurl{https://arxiv.org/abs/#1}}}

\bibitem[{Abadi {et~al.}(2016)Abadi, Barham, Chen, Chen, Davis, Dean, Devin,
  Ghemawat, Irving, Isard, {et~al.}}]{abadi2016tensorflow}
Abadi, M., Barham, P., Chen, J., {et~al.} 2016, in 12th $\{$USENIX$\}$
  Symposium on Operating Systems Design and Implementation ($\{$OSDI$\}$ 16),
  265--283

\bibitem[{Abbott {et~al.}(2016)Abbott, Abbott, Abbott, Abernathy, Acernese,
  Ackley, Adams, Adams, Addesso, Adhikari, Adya, Affeldt, Agathos, Agatsuma,
  Aggarwal, Aguiar, Aiello, Ain, Ajith, Allen, Allocca, Altin, Anderson,
  Anderson, Arai, Arain, Araya, Arceneaux, Areeda, Arnaud, Arun, Ascenzi,
  Ashton, Ast, Aston, Astone, Aufmuth, \& Aulbert}]{LIGO_2016}
Abbott, B.~P., Abbott, R., Abbott, T.~D., {et~al.} 2016, Phys. Rev. Lett., 116,
  061102, \dodoi{10.1103/PhysRevLett.116.061102}

\bibitem[{Ade {et~al.}(2016)Ade, Aghanim, Arnaud, Aumont, Baccigalupi, Banday,
  Barreiro, Bartolo, Battaner, Benabed, {et~al.}}]{Montier2016PlanckHighz}
Ade, P., Aghanim, N., Arnaud, M., {et~al.} 2016, A\&A, 596, A100,
  \dodoi{10.1051/0004-6361/201527206}

\bibitem[{Aghanim {et~al.}(2015)Aghanim, Altieri, Arnaud, Ashdown, Aumont,
  Baccigalupi, Banday, Barreiro, Bartolo, Battaner,
  {et~al.}}]{Aghanim2015Planck}
Aghanim, N., Altieri, B., Arnaud, M., {et~al.} 2015, A\&A, 582, A30,
  \dodoi{10.1051/0004-6361/201424790}

\bibitem[{Allen {et~al.}(2012)Allen, Anderson, Brady, Brown, \&
  Creighton}]{Allen2012}
Allen, B., Anderson, W.~G., Brady, P.~R., Brown, D.~A., \& Creighton, J. D.~E.
  2012, \prd, 85, \dodoi{10.1103/physrevd.85.122006}

\bibitem[{Andrews \& Phillips(2005)}]{Andrews2005}
Andrews, L.~C., \& Phillips, R.~L. 2005, Laser beam propagation through random
  media, Vol. 152 (SPIE press Bellingham, WA)

\bibitem[{Austermann {et~al.}(2018)Austermann, Beall, Bryan, Dober, Gao,
  Hilton, Hubmayr, Mauskopf, McKenney, Simon, {et~al.}}]{Austermann2018kids}
Austermann, J., Beall, J., Bryan, S., {et~al.} 2018, Journal of Low Temperature
  Physics, 193, 120, \dodoi{doi.org/10.1007/s10909-018-1949-5}

\bibitem[{Avila(2021)}]{Avila2021}
Avila, R. 2021, \mnras, \dodoi{/10.1093/mnrasl/slab080}

\bibitem[{Avila {et~al.}(2019)Avila, Vald{\'e}s-Hern{\'a}ndez, S{\'a}nchez,
  Cruz-Gonz{\'a}lez, Avil{\'e}s, Tapia-Rodr{\'\i}guez, \&
  Z{\'u}{\~n}iga}]{Avila2019SanPedro}
Avila, R., Vald{\'e}s-Hern{\'a}ndez, O., S{\'a}nchez, L., {et~al.} 2019,
  Monthly Notices of the Royal Astronomical Society, 490, 1397,
  \dodoi{10.1093/mnras/stz2672}

\bibitem[{Beland(1993)}]{Beland1993}
Beland, R.~R. 1993, Atmospheric Propagation of Radiation, 2, 157.
\newblock \url{https://apps.dtic.mil/dtic/tr/fulltext/u2/a364019.pdf#page=170}

\bibitem[{Bergstra {et~al.}(2011)Bergstra, Bardenet, Bengio, \&
  K{\'e}gl}]{bergstra2011algorithms}
Bergstra, J., Bardenet, R., Bengio, Y., \& K{\'e}gl, B. 2011, in 25th annual
  conference on neural information processing systems (NIPS 2011), Vol.~24,
  Neural Information Processing Systems Foundation.
\newblock
  \url{https://proceedings.neurips.cc/paper/2011/file/86e8f7ab32cfd12577bc2619bc635690-Paper.pdf}

\bibitem[{Blain {et~al.}(2002)Blain, Smail, Ivison, Kneib, \&
  Frayer}]{Blain2002}
Blain, A.~W., Smail, I., Ivison, R., Kneib, J.-P., \& Frayer, D.~T. 2002,
  \physrep, 369, 111, \dodoi{10.1016/S0370-1573(02)00134-5}

\bibitem[{Boyat \& Joshi(2015)}]{Boyat2015review}
Boyat, A.~K., \& Joshi, B.~K. 2015, arXiv preprint arXiv:1505.03489.
\newblock \url{https://arxiv.org/abs/1505.03489}

\bibitem[{Brien {et~al.}(2018)Brien, Ade, Barry, Castillo-Domìnguez, Ferrusca,
  Gascard, Gómez, Hargrave, Hornsby, Hughes, Pascale, Parrianen, Perez, Rowe,
  Tucker, González, \& Doyle}]{Brien2018}
Brien, T. L.~R., Ade, P. A.~R., Barry, P.~S., {et~al.} 2018, in Millimeter,
  Submillimeter, and Far-Infrared Detectors and Instrumentation for Astronomy
  IX, ed. J.~Zmuidzinas \& J.-R. Gao, Vol. 10708, International Society for
  Optics and Photonics (SPIE), 67 -- 75, \dodoi{10.1117/12.2313697}

\bibitem[{Bryan {et~al.}(2018)Bryan, Austermann, Ferrusca, Mauskopf, McMahon,
  Montaña, Simon, Novak, Sánchez-Argüelles, \& Wilson}]{Bryan2018}
Bryan, S., Austermann, J., Ferrusca, D., {et~al.} 2018, in Millimeter,
  Submillimeter, and Far-Infrared Detectors and Instrumentation for Astronomy
  IX, ed. J.~Zmuidzinas \& J.-R. Gao, Vol. 10708, International Society for
  Optics and Photonics (SPIE), 48 -- 55, \dodoi{10.1117/12.2314130}

\bibitem[{Casey {et~al.}(2014)Casey, Narayanan, \& Cooray}]{Casey2014}
Casey, C., Narayanan, D., \& Cooray, A. 2014, \physrep, 541, 45,
  \dodoi{10.1016/j.physrep.2014.02.009}

\bibitem[{Castillo-Dominguez {et~al.}(2018)Castillo-Dominguez, Ade, Barry,
  Brien, Doyle, Ferrusca, Gomez-Rivera, Hargrave, Hornsby, Hughes,
  {et~al.}}]{Castillo2018muscat}
Castillo-Dominguez, E., Ade, P., Barry, P., {et~al.} 2018, Journal of Low
  Temperature Physics, 193, 1010, \dodoi{10.1007/s10909-018-2018-9}

\bibitem[{Cho {et~al.}(2014)Cho, Van~Merri{\"e}nboer, Gulcehre, Bahdanau,
  Bougares, Schwenk, \& Bengio}]{cho2014learning}
Cho, K., Van~Merri{\"e}nboer, B., Gulcehre, C., {et~al.} 2014, arXiv preprint
  arXiv:1406.1078

\bibitem[{Church(1995)}]{Church1995}
Church, S. 1995, Monthly Notices of the Royal Astronomical Society, 272, 551,
  \dodoi{10.1093/mnras/272.3.551}

\bibitem[{Cuoco {et~al.}(2004)Cuoco, Cella, \& Guidi}]{Cuoco_2004}
Cuoco, E., Cella, G., \& Guidi, G.~M. 2004, Classical and Quantum Gravity, 21,
  S801, \dodoi{10.1088/0264-9381/21/5/061}

\bibitem[{DeNigris {et~al.}(2020)DeNigris, Wilson, Eiben, Lunde, Mauskopf, \&
  Contente}]{Denigris2020cryogenic}
DeNigris, N., Wilson, G., Eiben, M., {et~al.} 2020, Journal of Low Temperature
  Physics, 1, \dodoi{10.1007/s10909-019-02319-y}

\bibitem[{Downes {et~al.}(2012)Downes, Welch, Scott, Austermann, Wilson, \&
  Yun}]{Downes2012}
Downes, T., Welch, D., Scott, K., {et~al.} 2012, \mnras, 423, 529,
  \dodoi{10.1111/j.1365-2966.2012.20896.x}

\bibitem[{Errard {et~al.}(2015)Errard, Ade, Akiba, Arnold, Atlas, Baccigalupi,
  Barron, Boettger, Borrill, Chapman, {et~al.}}]{Errard2015modeling}
Errard, J., Ade, P., Akiba, Y., {et~al.} 2015, The Astrophysical Journal, 809,
  63, \dodoi{10.1088/0004-637X/809/1/63}

\bibitem[{George \& Huerta(2018)}]{George2018}
George, D., \& Huerta, E. 2018, Physical Review D, 97, 044039

\bibitem[{G{\'o}mez {et~al.}(2019)G{\'o}mez, Gonz{\'a}lez-Guti{\'e}rrez,
  Alonso, Santos, Rodr{\'\i}guez, Morris, Osborn, Basden, Bonavera,
  Gonz{\'a}lez, {et~al.}}]{Suarez2019experience}
G{\'o}mez, S. L.~S., Gonz{\'a}lez-Guti{\'e}rrez, C., Alonso, E.~D., {et~al.}
  2019, Publications of the Astronomical Society of the Pacific, 131, 108012

\bibitem[{Graves \& Schmidhuber(2005)}]{graves2005framewise}
Graves, A., \& Schmidhuber, J. 2005, Neural networks, 18, 602

\bibitem[{Greff {et~al.}(2017)Greff, Srivastava, Koutn{\'\i}k, Steunebrink, \&
  Schmidhuber}]{greff2017lstm}
Greff, K., Srivastava, R.~K., Koutn{\'\i}k, J., Steunebrink, B.~R., \&
  Schmidhuber, J. 2017, IEEE transactions on neural networks and learning
  systems, 28, 2222

\bibitem[{Hill {et~al.}(1980)Hill, Clifford, \& Lawrence}]{Hill1980refractive}
Hill, R., Clifford, S.~F., \& Lawrence, R.~S. 1980, JOSA, 70, 1192,
  \dodoi{https://doi.org/10.1364/JOSA.70.001192}

\bibitem[{Hochreiter {et~al.}(2001)Hochreiter, Bengio, Frasconi, Schmidhuber,
  {et~al.}}]{hochreiter2001gradient}
Hochreiter, S., Bengio, Y., Frasconi, P., Schmidhuber, J., {et~al.} 2001,
  Gradient flow in recurrent nets: the difficulty of learning long-term
  dependencies,  A field guide to dynamical recurrent neural networks. IEEE
  Press

\bibitem[{Hochreiter \& Schmidhuber(1997)}]{hochreiter1997long}
Hochreiter, S., \& Schmidhuber, J. 1997, Neural computation, 9, 1735

\bibitem[{Holland {et~al.}(2013)Holland, Bintley, Chapin, Chrysostomou, Davis,
  Dempsey, Duncan, Fich, Friberg, Halpern, {et~al.}}]{SCUBA2013}
Holland, W., Bintley, D., Chapin, E., {et~al.} 2013, \mnras, 430, 2513.
\newblock \url{https://doi.org/10.1093/mnras/sts612}

\bibitem[{Hughes {et~al.}(2010)Hughes, Correa, Schloerb, Erickson, Romero,
  Heyer, Reynoso, Narayanan, Perez-Grovas, Souccar, Wilson, \& Yun}]{LMT}
Hughes, D.~H., Correa, J.-C.~J., Schloerb, F.~P., {et~al.} 2010, in
  Ground-based and Airborne Telescopes III, ed. L.~M. Stepp, R.~Gilmozzi, \&
  H.~J. Hall, Vol. 7733, International Society for Optics and Photonics (SPIE),
  402 -- 414, \dodoi{10.1117/12.857974}

\bibitem[{Jamal \& Bloom(2020)}]{Jamal2020NN-VS}
Jamal, S., \& Bloom, J.~S. 2020, The Astrophysical Journal Supplement Series,
  250, 30, \dodoi{10.3847/1538-4365/aba8ff}

\bibitem[{Jordan(1997)}]{jordan1997serial}
Jordan, M.~I. 1997, in Advances in psychology, Vol. 121 (Elsevier), 471--495

\bibitem[{Kingma \& Ba(2014)}]{kingma2014adam}
Kingma, D.~P., \& Ba, J. 2014, arXiv preprint arXiv:1412.6980

\bibitem[{Kirchg{\"a}ssner {et~al.}(2012)Kirchg{\"a}ssner, Wolters, \&
  Hassler}]{Kirchgassner2012ModernTimeSeries}
Kirchg{\"a}ssner, G., Wolters, J., \& Hassler, U. 2012, Introduction to modern
  time series analysis (Springer Science \& Business Media)

\bibitem[{{Kolmogorov}(1991)}]{Kolmogorov1941}
{Kolmogorov}, A.~N. 1991, Proc. R. Soc. Lond. A, 434, 9,
  \dodoi{10.1098/rspa.1991.0075}

\bibitem[{Kov{\'a}cs(2008)}]{Kovacs2008crush}
Kov{\'a}cs, A. 2008, in Millimeter and Submillimeter Detectors and
  Instrumentation for Astronomy IV, Vol. 7020, International Society for Optics
  and Photonics, 70201S, \dodoi{10.1117/12.790276}

\bibitem[{Lay(1997)}]{Lay1997temporal}
Lay, O. 1997, Astronomy and Astrophysics Supplement Series, 122, 535,
  \dodoi{10.1051/aas:1997154}

\bibitem[{Lee \& Harp(1969)}]{Lee1969}
Lee, R.~W., \& Harp, J. 1969, Proc. IEEE, 57, 375,
  \dodoi{https://doi.org/10.1109/PROC.1969.6993}

\bibitem[{L{\"u}di {et~al.}(2005)L{\"u}di, Beyrich, \& M{\"a}tzler}]{Ludi2005}
L{\"u}di, A., Beyrich, F., \& M{\"a}tzler, C. 2005, Boundary-layer meteorology,
  117, 525, \dodoi{https://doi.org/10.1007/s10546-005-1751-1}

\bibitem[{Luo(2016)}]{luo2016review}
Luo, G. 2016, Network Modeling Analysis in Health Informatics and
  Bioinformatics, 5, 1

\bibitem[{Mairs {et~al.}(2015)Mairs, Johnstone, Kirk, Graves, Buckle, Beaulieu,
  Berry, Broekhoven-Fiene, Currie, Fich, {et~al.}}]{Mairs2015Comparison}
Mairs, S., Johnstone, D., Kirk, H., {et~al.} 2015, \mnras, 454, 2557,
  \dodoi{10.1093/mnras/stv2192}

\bibitem[{Martinache {et~al.}(2018)Martinache, Rettura, Dole, Lehnert, Frye,
  Altieri, Beelen, B{\'e}thermin, Le~Floc’h, Giard,
  {et~al.}}]{Martinache2018SPHerIC}
Martinache, C., Rettura, A., Dole, H., {et~al.} 2018, A\&A, 620, A198,
  \dodoi{10.1051/0004-6361/201833198}

\bibitem[{Marulanda {et~al.}(2020)Marulanda, Santa, \&
  Romano}]{marulanda2020deep}
Marulanda, J.~P., Santa, C., \& Romano, A.~E. 2020, Physics Letters B, 810,
  135790

\bibitem[{McCloskey \& Cohen(1989)}]{mccloskey1989catastrophic}
McCloskey, M., \& Cohen, N.~J. 1989, in Psyfchology of learning and motivation,
  Vol.~24 (Elsevier), 109--165

\bibitem[{Milotti(2002)}]{Milotti2002}
Milotti, E. 2002, arXiv preprint physics/0204033.
\newblock \url{https://arxiv.org/abs/physics/0204033}

\bibitem[{{Monta{\~n}a} {et~al.}(2019){Monta{\~n}a}, {Ch{\'a}vez Dagostino},
  {Aretxaga}, {Novak}, {Pope}, {Wilson}, \& {TolTEC Team}}]{Montana}
{Monta{\~n}a}, A., {Ch{\'a}vez Dagostino}, M., {Aretxaga}, I., {et~al.} 2019,
  \memsai, 90, 632

\bibitem[{Nieuwstadt {et~al.}(2016)Nieuwstadt, Westerweel, \&
  Boersma}]{Nieuwstadt2016turbulence}
Nieuwstadt, F.~T., Westerweel, J., \& Boersma, B.~J. 2016, Turbulence:
  introduction to theory and applications of turbulent flows (Springer)

\bibitem[{Obhukov(1941)}]{Obhukov1941}
Obhukov, A. 1941, Izv. Akad. Nauk SSSR, Ser. Geofiz, 5, 453

\bibitem[{Osborn(2015)}]{Osborn2015dishcorrection}
Osborn, J. 2015, Monthly Notices of the Royal Astronomical Society, 446, 1305,
  \dodoi{10.1093/mnras/stu2175}

\bibitem[{Osborn {et~al.}(2016)Osborn, Butterley, Townson, Reeves, Morris, \&
  Wilson}]{Osborn2016C2n}
Osborn, J., Butterley, T., Townson, M., {et~al.} 2016, Monthly Notices of the
  Royal Astronomical Society, stw2685, \dodoi{10.1093/mnras/stw2685}

\bibitem[{Osborn {et~al.}(2015)Osborn, F{\"o}hring, Dhillon, \&
  Wilson}]{Osborn2015origins}
Osborn, J., F{\"o}hring, D., Dhillon, V., \& Wilson, R. 2015, Monthly Notices
  of the Royal Astronomical Society, 452, 1707, \dodoi{10.1093/mnras/stv1400}

\bibitem[{Pan \& Yang(2009)}]{pan2009survey}
Pan, S.~J., \& Yang, Q. 2009, IEEE Transactions on knowledge and data
  engineering, 22, 1345

\bibitem[{Pascanu {et~al.}(2013)Pascanu, Mikolov, \&
  Bengio}]{pascanu2013difficulty}
Pascanu, R., Mikolov, T., \& Bengio, Y. 2013, in International conference on
  machine learning, 1310--1318

\bibitem[{Pope(2000)}]{Pope2000}
Pope, S.~B. 2000, Turbulent Flows (Cambridge University Press),
  \dodoi{https://doi.org/10.1017/CBO9780511840531}

\bibitem[{Qian {et~al.}(2018)Qian, Yao, Wang, Wang, Bai, \&
  Yin}]{Qian2018AliObs}
Qian, X., Yao, Y., Wang, H., {et~al.} 2018, Publications of the Astronomical
  Society of the Pacific, 130, 125002, \dodoi{10.1088/1538-3873/aae6e2}

\bibitem[{Rae {et~al.}(2019)Rae, Potapenko, Jayakumar, \&
  Lillicrap}]{rae2019compressive}
Rae, J.~W., Potapenko, A., Jayakumar, S.~M., \& Lillicrap, T.~P. 2019, arXiv
  preprint arXiv:1911.05507

\bibitem[{Ratcliff(1990)}]{ratcliff1990connectionist}
Ratcliff, R. 1990, Psychological review, 97, 285

\bibitem[{Robey {et~al.}(1992)Robey, Fuhrmann, Kelly, \&
  Nitzberg}]{Robey92_AMF}
Robey, F., Fuhrmann, D., Kelly, E., \& Nitzberg, R. 1992, IEEE Transactions on
  Aerospace and Electronic Systems, 28, 208, \dodoi{10.1109/7.135446}

\bibitem[{Rodr{\'\i}guez-Montoya {et~al.}(2018)Rodr{\'\i}guez-Montoya,
  S{\'a}nchez-Arg{\"u}elles, Aretxaga, Bertone, Ch{\'a}vez-Dagostino, Hughes,
  Monta{\~n}a, Wilson, \& Zeballos}]{Rodriguez2018}
Rodr{\'\i}guez-Montoya, I., S{\'a}nchez-Arg{\"u}elles, D., Aretxaga, I.,
  {et~al.} 2018, ApJS, 235, 12, \dodoi{10.3847/1538-4365/aaa83c}

\bibitem[{Sayers {et~al.}(2010)Sayers, Golwala, Ade, Aguirre, Bock, Edgington,
  Glenn, Goldin, Haig, Lange, Laurent, Mauskopf, Nguyen, Rossinot, \&
  Schlaerth}]{Sayers2010}
Sayers, J., Golwala, S.~R., Ade, P.~A.~R., {et~al.} 2010, ApJ, 708, 1674.
\newblock \url{http://stacks.iop.org/0004-637X/708/i=2/a=1674}

\bibitem[{Schuster \& Paliwal(1997)}]{schuster1997bidirectional}
Schuster, M., \& Paliwal, K.~K. 1997, IEEE Transactions on Signal Processing,
  45, 2673

\bibitem[{Scott {et~al.}(2008)Scott, Austermann, Perera, Wilson, Aretxaga,
  Bock, Hughes, Kang, Kim, Mauskopf, {et~al.}}]{Scott2008pipeline}
Scott, K., Austermann, J., Perera, T.~A., {et~al.} 2008, Monthly Notices of the
  Royal Astronomical Society, 385, 2225

\bibitem[{Solignac {et~al.}(2012)Solignac, Brut, Selves, B{\'e}teille, \&
  Gastellu-Etchegorry}]{Solignac2012}
Solignac, P.~A., Brut, A., Selves, J.-L., B{\'e}teille, J.-P., \&
  Gastellu-Etchegorry, J.-P. 2012, Boundary-layer meteorology, 143, 261,
  \dodoi{https://doi.org/10.1007/s10546-011-9692-3}

\bibitem[{{Tatarskii}(1971)}]{Tatarskii1971}
{Tatarskii}, V.~I. 1971, {The effects of the turbulent atmosphere on wave
  propagation} (Jerusalem: Israel Program for Scientific Translations, 1971).
\newblock \url{https://ui.adsabs.harvard.edu/abs/1971etaw.book.....T}

\bibitem[{Tatarskii(2016)}]{Tatarskii2016wave}
Tatarskii, V.~I. 2016, Wave propagation in a turbulent medium (Courier Dover
  Publications).
\newblock \url{https://lccn.loc.gov/2016028015}

\bibitem[{Taylor(1938)}]{Taylor1938}
Taylor, G.~I. 1938, Proceedings of the Royal Society of London. Series
  A-Mathematical and Physical Sciences, 164, 476,
  \dodoi{https://doi.org/10.1098/rspa.1938.0032}

\bibitem[{Turin(1960)}]{Turin1960MF}
Turin, G. 1960, IRE Transactions on Information Theory, 6, 311,
  \dodoi{10.1109/TIT.1960.1057571}

\bibitem[{Vallis(2017)}]{Vallis2017atmospheric}
Vallis, G.~K. 2017, Atmospheric and oceanic fluid dynamics (Cambridge
  University Press)

\bibitem[{Vaswani {et~al.}(2017)Vaswani, Shazeer, Parmar, Uszkoreit, Jones,
  Gomez, Kaiser, \& Polosukhin}]{vaswani2017attention}
Vaswani, A., Shazeer, N., Parmar, N., {et~al.} 2017, in Advances in neural
  information processing systems, 5998--6008

\bibitem[{Von~Karman(1948)}]{vonKarman1948}
Von~Karman, T. 1948, Proceedings of the National Academy of Sciences of the
  United States of America, 34, 530,
  \dodoi{https://doi.org/10.1073/pnas.34.11.530}

\bibitem[{Wang {et~al.}(2021)Wang, Azam, Wilson, Neville, \&
  Morina}]{Wang2021generating}
Wang, Y., Azam, A., Wilson, M.~C., Neville, A., \& Morina, A. 2021, Proceedings
  of the Institution of Mechanical Engineers, Part J: Journal of Engineering
  Tribology, 235, 2640, \dodoi{10.1177/13506501211049624}

\bibitem[{Wang {et~al.}(2018)Wang, Liu, Zhang, \& Wang}]{Wang2018simulation}
Wang, Y., Liu, Y., Zhang, G., \& Wang, Y. 2018, Journal of Tribology, 140,
  021403, \dodoi{10.1115/1.4037793}

\bibitem[{Weiss {et~al.}(2016)Weiss, Khoshgoftaar, \& Wang}]{weiss2016survey}
Weiss, K., Khoshgoftaar, T.~M., \& Wang, D. 2016, Journal of Big data, 3, 9

\bibitem[{Wheelon(2001)}]{Wheelon2001}
Wheelon, A.~D. 2001, Electromagnetic Scintillation: Volume 1, Geometrical
  Optics (Cambridge University Press).
\newblock \url{http://www.cambridge.org/9780521801980}

\bibitem[{Wheelon(2003)}]{Wheelon2003}
---. 2003, Electromagnetic scintillation: volume 2, weak scattering (Cambridge
  University Press).
\newblock \url{http://www.cambridge.org/9780521801997}

\bibitem[{Wilson {et~al.}(2008)Wilson, Austermann, Perera, Scott, Ade, Bock,
  Glenn, Golwala, Kim, Kang, {et~al.}}]{wilson2008aztec}
Wilson, G., Austermann, J., Perera, T., {et~al.} 2008, Monthly Notices of the
  Royal Astronomical Society, 386, 807

\bibitem[{Yuan {et~al.}(2015)Yuan, Luo, Sun, Zeng, Ge, \& Fu}]{Yuan2015}
Yuan, R., Luo, T., Sun, J., {et~al.} 2015, Atmospheric Chemistry and Physics,
  15, 2521, \dodoi{https://doi.org/10.5194/acp-15-2521-2015}

\bibitem[{Zaremba \& Sutskever(2014)}]{learningToExecute}
Zaremba, W., \& Sutskever, I. 2014, International conference on learning
  representations

\bibitem[{Zavala {et~al.}(2018)Zavala, Monta{\~n}a, Hughes, Yun, Ivison,
  Valiante, Wilner, Spilker, Aretxaga, Eales, {et~al.}}]{Zavala2018}
Zavala, J.~A., Monta{\~n}a, A., Hughes, D.~H., {et~al.} 2018, Nature Astronomy,
  2, 56, \dodoi{10.1038/s41550-017-0297-8}

\bibitem[{Zhang {et~al.}(2018)Zhang, Geiger, Pohjalainen, Mousa, Jin, \&
  Schuller}]{Zhang2018review}
Zhang, Z., Geiger, J., Pohjalainen, J., {et~al.} 2018, ACM Transactions on
  Intelligent Systems and Technology (TIST), 9, 1, \dodoi{10.1145/3178115}

\end{thebibliography}
\addcontentsline{toc}{section}{References}

\end{document}